# Investigation of the Short Argon Arc with Hot Anode, Part II: Analytical Model

A. Khrabry[1], I.D. Kaganovich[1], V. Nemchinsky[2], A. Khodak[1]

[1]*Princeton Plasma Physics Laboratory, Princeton NJ, 08543 USA*

[2]*Keiser University, Fort Lauderdale, FL, 33309 USA*

**Abstract**

Short atmospheric pressure argon arc is studied numerically and analytically. In a short arc with inter-electrode gap of several millimeters non-equilibrium effects in plasma play important role in operation of the arc. High anode temperature leads to electron emission and intensive radiation from its surface. Complete self-consistent analytical model of the whole arc comprising of models for near-electrode regions, arc column and a model of heat transfer in cylindrical electrodes was developed. The model predicts width of non-equilibrium layers and arc column, voltages and plasma profiles in these regions, heat and ion fluxes to the electrodes. Parametric studies of the arc have been performed for a range of the arc current densities, inter-electrode gap widths and gas pressures. The model was validated against experimental data and verified by comparison with numerical solution. Good agreement between the analytical model and simulations and reasonable agreement with experimental data were obtained.

## I. Introduction

Atmospheric pressure arcs recently found application in production of nanoparticles, such as carbon nanotubes[1,2,3,4,5] and boron-nitride nanotubes[6]. Distinguishing features of such arcs are typically short length of about several millimeters between electrodes and hot ablating anode characterized by intensive electron emission and radiation from its surface. Electrode ablation significantly increases complexity of the arc physics and chemistry. As a first step a short argon arc with cylindrical tungsten electrodes is studied in this series of papers. No ablation takes place from tungsten electrodes but effects of emission, radiation, non-equilibrium layers are still pronounced. Results for carbon arc with graphite electrodes in helium atmosphere will be presented in subsequent publications.

The first paper of the series[7] was dedicated to numerical simulation of an argon arc with cylindrical tungsten electrodes with emphasis on non-equilibrium effects in the near-electrode regions. It was shown that the non-equilibrium effects play important role in operation of the arc and should be taken into account in modeling. It was also shown that the electron emission from the anode surface can significantly affect the potential drop in the plasma region near the anode.

Though the numerical simulations can yield all arc plasma profiles, theoretical analysis can unravel complicated physical processes underpinning the arc self-organization. Scaling laws of crucial arc characteristics such as potential drops in different arc regions, heat fluxes to electrodes etc. with arc current, pressure and inter-electrode width obtained theoretically are used for planning experimental campaigns and choosing right arc parameters for arc applications. Self-consistent analytical model of the whole arc is helpful for fast assessments of arc experimental setup design. These predictions are also important for verification of numerical codes, for instance, for verifying whether results of the numerical simulations exhibit correct asymptotical behavior. Developed analytical model provides understanding of which terms in the governing equations are of major importance and which can be neglected. This knowledge can be used for simplification of numerical codes.

Argon arc was extensively studied previously. However, among modeling papers mostly numerical studies rather than analytical studies are present in literature. Significant part of the studies is focusing on one or another part of the arc, not considering arc as a whole. For example, numerous numerical studies of the anodic region can be found in literature[8,9,10,11,12]. Thorough reviews on numerical and experimental studies of the near-anode region of arc discharges can be found in Refs. [13 and 14]. Approximate relation for heat flux to the anode, heuristic assessment for the width of the near-anode region can be found in Ref. [14], for instance. However, we could not find self-consistent analytical model of the region, providing accurate relations for its width and voltage. We also could not find papers considering hot electron emitting anode. Analytical models of the arc column are limited to the case of long arcs where no variation of plasma parameters along the axis takes place in the arc column[15,16]. Analytical studies of the cathodic region either focus on energy balance[17,18,19] or on the ion transport[20,21]. We could not find self-consistent model of the cathodic region coupling all the effects of heat conduction in the electrode, ion generation and transport, and the sheath voltage drop.

The 1D model of argon arc presented in the first paper of the series[7] features non-equilibrium plasma transport equations with the transport coefficients derived from kinetic theory[22]. The results of numerical solution of the non-equilibrium plasma transport equations were compared with simulations of Ref. [22] and validated against experimental data[23,24]. The transport equations were thoroughly described in Ref. [7] and will be used in the current paper.

Parametric studies of the atmospheric pressure argon arc for various current densities and inter-electrode gap sizes were performed in Ref. [7]. It was shown, in particular, that the different arc regions are rather autonomic even in case of short arcs (weakly depend on the arc length) and can be considered separately. Based on the results of the simulations performed and presented in the first paper[7], the current study reports self-consistent analytic models of the near-electrode regions and the arc column combined into a unified self-consistent analytical model of the whole arc. Non-equilibrium processes in plasma and effects of near-electrode space-charge sheaths are taken into account. The analytical model is capable to predict the arc structure, plasma parameters and voltages in different arc regions, their sizes, and heat fluxes to the electrodes. The analytical arc model was benchmarked against the simulations and validated against experimental data of Ref. [23]. Results for background argon pressures of 1 atm. and 3 atm. will be presented.

The organization of the paper is as follows. In Section II, system of equations describing non-equilibrium transport processes in the arc are given. Model of the cathodic region is presented in section III providing relations for the region width, voltage, heat flux and ion current to the cathode and electron temperature. Section IV is devoted to the arc column, where it is shown that a single differential equation for the gas temperature profile can describe the arc column; asymptotic solutions for the temperature profile and relation for the arc column voltage are derived. In section V, analytic model of the anodic region is given providing relations for the anodic region width, voltage and heat flux to the electrode. In section VI, developed asymptotic solutions for the all arc regions are used to calculate VAC of the entire arc. Theoretical results for VAC are also validated against the available experimental data. Conclusions of this work are summarized in Section VII.

## II. Basic transport equations in the arc model

In this section, system of transport equations describing species transport and heat transfer in the arc is presented. Full set of governing transport equations for the arc was already given in the first paper of this series[7] and in Ref. [22] with proper description and derivation. Equations are formulated for quasi-neutral plasma outside space-charge sheaths. Here, we repeat these equations briefly.

Electric field can be expressed using generalized Ohm's law that takes into account electron diffusion, thermal diffusion and electron-ion friction:

$$\vec{E} = -\frac{k}{e}\left(1 + C_e^{(e)}\right)\nabla T_e - \frac{k}{e}T_e\frac{\nabla n}{n} - \frac{\vec{\Gamma}_e}{ne}m_e\left(\nu_{e,a} + \nu_{e,i}\right) + \frac{m_e}{e}\left(\nu_{e,a}\frac{\vec{\Gamma}_a}{n_a} + \nu_{e,i}\frac{\vec{\Gamma}_i}{n}\right), \tag{1}$$

where $e$ is elementary charge, $k$ is the Boltzmann constant, $n \equiv n_e = n_i$ is density of electrons and ions (quasi-neutrality approximation is used in this paper), $n_a$ is density of neutral atoms and other variables are defined below.

For the ion transport we use ion continuity equation, where the electric field is excluded using electron flux, $\vec{\Gamma}_e$, and assuming that gas velocity is negligible in 1D approximation. This gives for the ion flux, $\vec{\Gamma}_i$, the ambipolar diffusion and thermal diffusion (see Ref. [7], equation (39)):

$$\vec{\Gamma}_i = -D\nabla n - n(D_T \nabla \ln T + D_{Te} \nabla \ln T_e) - A_e \vec{\Gamma}_e, \tag{2}$$

where:

$D = k(T + T_e)/(0.5(\nu_{i,a} + \nu_{a,i})m_{Ar} + \nu_{a,e}m_e)$ is the ambipolar diffusion coefficient,

$D_T = DT/(T + T_e)$, $D_{Te} = DT_e/(T + T_e)$ is the thermal diffusion coefficient,

$A_e = (\nu_{e,a}m_e)/(0.5(\nu_{i,a} + \nu_{a,i})m_{Ar} + \nu_{a,e}m_e)$ is a kinetic coefficient,

$\nu_{k,j}$ is the effective collision frequency of species *k* with species *j*,

$$v_{k,j} = \frac{4}{3}\sqrt{\frac{8kT_{kj}}{\pi m_{kj}}} C_{kj} \sigma_{kj} n_j,$$

subscripts $k, j$ denote different species: argon atoms $a$, argon ions $i$ and electrons e, $n_k$ – number density of species $k$,

$$m_{kj} = \frac{m_k m_j}{m_k + m_j}, \quad T_{kj} = \frac{m_k T_j + m_j T_k}{m_k + m_j}.$$

Here, $m_k$ is mass of particles of a sort $k$, $T_k$ is their temperature, $\vec{\Gamma}_k$ is flux of species $k$, $\sigma_{kj}$ is collision cross-section. Temperatures and masses of heavy particles are very close and are not distinguished in the model: $T_i = T_a = T$, $m_i = m_a = m_{Ar}$.

For electron-ion collisions cross-section is:

$$\sigma_{ei} = \frac{e^4 \ln \Lambda}{32\pi\varepsilon_0^2 (kT_e)^2}, \tag{3}$$

where $\varepsilon_0$ is vacuum permittivity, $\ln \Lambda = \ln\left(8\pi\varepsilon_0 kT_e \sqrt{\varepsilon_0 kT_e/n}/e^3\right)$ is Coulomb logarithm, $C_{kj}$, $C_k^{(e)}$ are numerical coefficients of order of unity that are given in Refs. [7, 22]. For strongly-ionized plasma, $C_e^{(e)} = 0.7$.

In quasineutral approximation, electron and ion density is determined by continuity equation (see Ref. [7], equation (40)):

$$\nabla\left(n\vec{V}_T - D\nabla n\right) = s_i + \vec{\Gamma}_e \cdot \nabla A_e, \tag{4}$$

where $\vec{V}_T \equiv D_T \nabla \ln T + D_{T_e} \nabla \ln T_e$ accounts for thermal diffusion effects, $s_i = k_i n n_a - k_r n^3$ is volumetric plasma source (ionization) and sink (three body recombination), $k_i$, $k_r$ are the temperature-dependent reaction rate coefficients. Arrhenius-like approximations for $k_i$ and $k_r$ were used in the analytical model:

$$k_i(T) = A_i \exp(-T_i/T), \quad k_r(T) = A_r \exp(T_r/T), \tag{5}$$

$A_i = 1.5 \cdot 10^{-14} m^3/s$, $T_i = 140\,000 K$, $A_r = 10^{-43} m^6/s$, $T_r = 51\,000 K$.

These simple relations are rather good approximations of more accurate formulae, in which $k_i$ is calculated as described in Ref. [25], and $k_r$ is calculated to satisfy the ionization-recombination balance (deviation between accurate and approximate values does not exceed 20% in a temperature range 5000 K – 16000 K):

$$k_i n_a = k_r n_{Saha}^2,  \qquad (6)$$

where $n_{e,Saha}$ is the equilibrium number density defined by the Saha equation:

$$\frac{n_{Saha}^2}{n_a} = 2 g_i / g_a \left(\frac{2\pi m_e k T_e}{h^2}\right)^{3/2} \exp\left(-\frac{eE_{ion}}{kT_e}\right). \qquad (7)$$

Here, $g_i / g_a = 6$ is a ratio of statistical weights of ground state and ionized state, see Ref. [26], $h$ is Planck's constant, $E_{ion}$ is the ionization energy of argon atoms.

Note that similar approximations of coefficients (5) can be found in other papers[27,28].

Transport of energy of electrons and heavy particles are described by following equations:

$$\nabla \cdot \left(3.2 \frac{k}{e} T_e \vec{\Gamma}_e\right) = \nabla \cdot (\lambda_e \nabla T_e) - e\vec{\Gamma}_e \cdot \vec{E} - Q^{e-h} - Q^{ion} - Q^{rad}, \qquad (8)$$

$$0 = \nabla \cdot (\lambda_h \nabla T) + Q^{e-h} + \vec{\Gamma}_i \cdot \vec{E}, \qquad (9)$$

where coefficient 3.2 in the left-hand side of Eq. (8) is derived from kinetic theory in the limit of strongly ionized plasma ($\nu_{e,a} \ll \nu_{e,i}$), the electron-ion collision frequency is large compared to electron-atom collision frequency ($3.2 = 2.5 + A_i^{(e)} + A_a^{(e)}$, see Refs. [7, 22] for details). As was shown in Ref. [7], this condition is valid for most of the arc including major part of the near-electrode non-equilibrium regions. $\lambda_e$ is the thermal conductivity of electron gas[22]:

$$\lambda_e = kn \frac{3\pi}{10} \sqrt{\frac{8kT_e}{\pi m_e}} \frac{1}{\sigma_{ei}},$$

$\lambda_h$ is the thermal conductivity of and heavy particles, $Q^{e-h} = A^{e-H}(T_e - T)$ is the volumetric heat exchange between electrons and heavy particles,

$$A^{e-H} = 8 n_e^2 \sigma_{ei} \sqrt{\frac{2km_e T_e}{\pi}} \frac{k}{m},$$

$Q^{rad}$ represents the volumetric radiation losses[22]:

$$Q^{rad} = 2.6 \times 10^{25} W/m^3 \frac{p}{1 atm} \left(\frac{1K}{T_e}\right)^{2.52} \exp\left(-\frac{1.69 \times 10^5 K}{T_e}\right). \tag{10}$$

Equation (8) is a simplified version of equation (12) of Ref. [7], where it was taken into account that the ion current is small compared to the arc current everywhere except for the near-cathode region where ionization degree is small.

In 1D approximation there is no gas flow, and the total plasma and gas pressure is constant:

$$nkT_e + (n_a + n)kT = p. \tag{11}$$

Electrode temperatures are important parameters that affect current propagation due to electron emission. To determine electrode temperatures the heat transfer equations have to be solved in electrodes. Heat transfer equations along the axis of cylindrical electrodes can be reduced to 1D approximation by neglecting temperature variation in radial direction and temperature profile is governed by the following equation:

$$\pi r_{el}^2 \frac{d}{dx}\left(\lambda_{el}\frac{dT}{dx}\right) = 2\pi\left[(T-T_{amb})\lambda_{gas} Nu + \varepsilon\sigma(T^4-T_{amb}^4)\right] + \pi r_{el}^2 j^2 \rho_{el}. \tag{12}$$

Here, $\lambda_{el}$ is thermal conductivity of the electrode material (assumed to be constant, 170 W/m/K for tungsten), $r_{el}$ is the electrode radius, $T_{amb} = 300K$ is the ambient temperature, $\lambda_{gas} \cong 0.1 W/m/K$ is the thermal conductivity of gas surrounding the electrode, $Nu$ is the Nusselt number taken equal to 1.1, see Ref. [29]; $\sigma$ is the Stefan-Boltzmann constant, $\varepsilon$ is the emissivity taken equal to 1, $\rho_{el}$ is the electrical resistivity of the electrode material (assumed to be constant, small for metallic electrodes), $j$ is the current density, assumed constant along the arc.

At the plasma facing surface of the electrode, the heat flux from plasma can be used as a boundary condition:

$$q_{el,tip} = \lambda_{el}\left(\frac{dT}{dx}\right)_{front} = q_{to\ electrode} - q_{rad,front}. \tag{13}$$

Here, $q_{to\ electrode}$ denotes the heat flux from plasma to the electrode, $q_{rad,front}$ denotes the radiation heat flux from the front surfaces of the electrodes including mutual radiation[29]. However, according to results of the 1D simulations performed in Ref. [7], the net radiation from the front surfaces of thin electrodes appeared to be of minor importance (due to significant portion of incident radiation from the opposite electrode) and is not taken into account in analytic model. At the opposite surface of the electrode (away from plasma), one can use the ambient gas temperature condition. If the electrode is sufficiently long, then all the heat from plasma and Joule heat generated inside the electrode are lost at side walls due to radiation and thermal conduction into the ambient gas. A condition of vanishing heat flux at the opposite surface of the electrode (away from plasma) can be used in this case. With this boundary condition and for constant transport coefficients equation (12) can be also solved analytically to yield:

$$q_{h.cond.}(T) = T\sqrt{2\frac{\lambda_{el}}{r_{el}}\left(Nu\frac{\lambda_{gas}}{r_{el}}\left(1-2\frac{T_{amb}}{T}\right)+\frac{2}{5}\sigma\varepsilon\left(T^3-\frac{T_{amb}^4}{T}\right)\right)-2\frac{j^2\rho_{el}\lambda_{el}}{T}}, \quad (14)$$

where $q_{h.cond.}$ is the heat flux through a cross-section with temperature $T$; $T_{amb}$ is temperature of the ambient gas. Solution (14) is used in the analytical model described further.

Substitution of temperature at a front surface of an electrode $T_{el}$ in relation (14) yields heat flux into the electrode from the plasma. Because the electrode temperature is much higher than ambient temperature and electrical resistivity of metallic electrodes is negligible; relation (14) can be significantly simplified, and the heat flux into the electrode can be expressed by:

$$q_{h.cond.} \cong T_{el}^{2.5}\sqrt{\frac{4}{5}\frac{\lambda_{el}}{r_{el}}\sigma\varepsilon}. \quad (15)$$

In (15) it was assumed that the electrode radius is not less than 1 mm in order to neglect the term accounting for thermal conduction of the ambient gas. Solution (15) is used in the analytical model described further.

## III. Model of the cathodic region

### III.1. Voltage in the near-cathode layer

Simulation results for the near-cathode region of atmospheric pressure arc in 1D approximation are shown in figure 1 for various current densities. In the simulations, electrode temperature was determined from the self-consistent heat transfer equations between plasma and the cathode; the electrode diameter is 6mm.

Deviation from the ionization and thermal equilibrium is clearly evident in figures 1(a) and 1(b). In the plasma bulk, temperatures of electrons and heavy particles are equal due to high collisional heat exchange, plasma density can be described by equilibrium relation (the Saha equation) (7). The electron temperature at the cathode is high due to high energy of the emitted electrons after they have been accelerated inside the cathode sheath. While moving inside the plasma, these electrons lose their energy due to exchange with colder bulk plasma electrons and due to inelastic processes (ionization and excitation). Accordingly, the electron temperature decreases towards the plasma, whereas the temperature of heavy particles decreases towards the cathode and becomes equal to the electrode temperature at the cathode front surface. Elevated electron temperature implies increase of the equilibrium electron number density according to the Saha equation (shown by dotted lines in the Fig. 1(b) for ionization equilibrium, $n_{e,Saha}$), whereas the actual plasma density decreases due to ion acceleration towards the cathode surface. Difference between equilibrium and actual plasma densities results in high net production of ions that move towards the cathode due to the drift in the electric field and diffusion driven by the plasma density gradient.

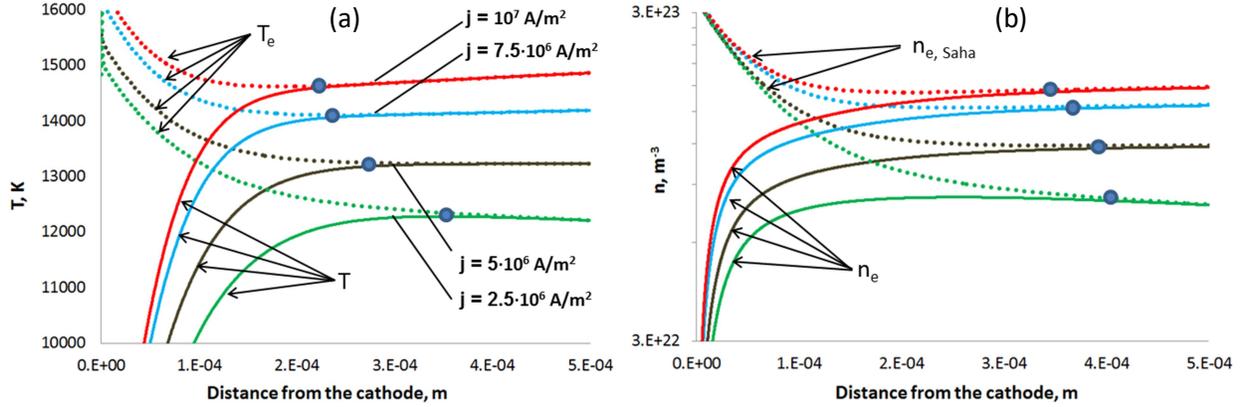

**Figure 1. Results of the 1D simulations for the near-cathode region of atmospheric pressure arc: (a) electron (dashed) and gas temperatures; (b) equilibrium and actual electron density.**

In order to determine voltage drop in the near-cathode region of plasma it is convenient to consider energy balance in this region (see figure 2), as it was done, for example, in Refs. [17, 18, and 19]. Consider integral energy balance in the region. Energy released in the near-cathode region is a product of the current density and voltage drop. This energy is transferred to bulk plasma and to the cathode. As shown in Ref. [7], the heat flux in plasma outside the near-cathode region is mostly transferred by convection of electrons; that is contribution of the thermal conductivity can be neglected. The electron current constitutes most of the total current density (the ion current is small compared to the total current). Accordingly, the heat flux from the near-cathode region to the plasma can be written as $3.2(k/e)jT_{e,plasma}$. The simulations have shown that this simplification is valid for rather short arcs until near-electrode regions start to overlap. Note that in Refs. [17 and 18] the heat flux to plasma was neglected for simplicity, and in Ref. [19] similar simplifications to those described above were used.

Composition of energy flux from plasma to the cathode is not important for this derivation. It will be considered further in this section. Resultant energy balance relation for the near-cathode region reads:

$$jV_{c\,layer} = q_{to\,cathode} + 3.2\frac{k}{e}jT_{e,plasma}. \tag{16}$$

Consider the heat balance at the cathode surface: the heat flux to the cathode from plasma (see figure 2) is partially spent on electrode cooling by electron emission; because emitted electrons overcome the surface potential barrier, i.e. work function when exiting the electrode. This heat flux is equal to $jV_w$.

Both thermionically emitted electrons and electrons neutralizing the ion flux are included giving the total current as a sum. The rest of the heat is transferred into the cathode body by heat conduction; heat radiation from the cathode front surface is small compared to $jV_w$ and is neglected:

$$q_{to\ cathode} = jV_w + q_{c.h.cond.} \qquad (17)$$

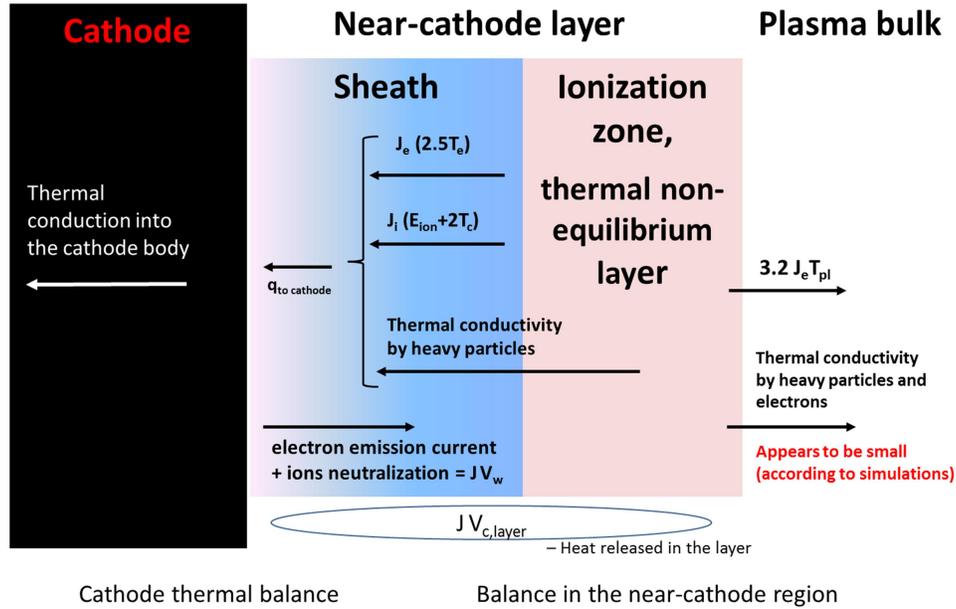

**Figure 2. Schematic energy balance in the cathodic region of the arc.**

Substitution of Eq. (16) into Eq. (17) gives cathode layer voltage drop:

$$V_{c\ layer} = V_w + 3.2 \frac{k}{e} T_{e,plasma} + \frac{q_{c.h.cond.}}{j}. \qquad (18)$$

Unknowns here are $T_{e,plasma}$ and $q_{c.h.cond.}$. A term with $T_{e,plasma}$ in (18) representing convection appears to be of order of work function and cannot be omitted. The value of $T_{e,plasma}$ can be determined from ion balance in the near-cathode plasma region, as described in section III.3, equation (29). Note that, 1D simulations show that $T_{e,plasma}$ does not change significantly with the arc current or pressure. It varies in a range from about 12 000 K to 16 000 K with current density variation form $2 \times 10^6$ A/m$^2$ to $2 \times 10^7$ A/m$^2$. For the sake of simplicity, approximate constant value of 14 000 K can be used for current densities in the range considered, variation of $T_{e,plasma}$ around 14 000 K gives an error not exceeding 0.5 V (which is less than 5% for the conditions considered). For better accuracy, or for different arc operating conditions, one can obtain $T_{e,plasma}$ from equation (29). The results presented below were obtained with the constant value of $T_{e,plasma}$ equal to 14 000 K.

Conductive heat flux into the cathode $q_{c.h.cond.}$ used in Eqs. (18) and (29) can be determined by substitution of the electrode front surface temperature $T_c$ into Eq. (15) yielding following relation for the cathode voltage:

$$V_{c\,layer} = V_w + 3.2 \frac{k}{e} \cdot T_{e,plasma} + T_c^{2.5} \frac{1}{j} \sqrt{\frac{4}{5} \frac{\lambda_c}{r_c} \sigma \varepsilon} \ . \tag{19}$$

Here, $\lambda_c$ is thermal conductivity of the cathode material, $r_c$ is the cathode radius. Note that in the papers [17, 18 and 19] wide and long cathodes were considered, radiation from the cathode surface was neglected resulting in a simpler relation for the heat flux into the cathode compared to Eq. (15) for the thin cylindrical cathode with radiation.

Temperature of the cathode front surface $T_c$ can be determined from current conservation at the cathode surface:

$$j = j_R + j_{i,c} - j_e^{plasma} \ . \tag{20}$$

Here, $j_{i,c}$ is ion current at the cathode surface, $j_e^{plasma}$ is current of plasma electrons to the cathode, which typically negligible due to its suppression by the cathode sheath voltage drop, $j_R$ is emission current described by the Richardson formula:

$$j_R(T_{el}) = A_R T_{el}^2 \exp\left(-\frac{e(V_w + E_{Schott})}{kT_{el}}\right) . \tag{21}$$

Here, $A_R$ is Richardson's constant, $V_w$ is the work function of the electrode material (4.5 V for tungsten), $E_{Schott}$ is the Schottky correction voltage (about 0.1 V, see Ref. [20] for instance).

Substitution of (21) into (20) and neglecting the Shottky voltage yields a relation for the cathode surface temperature $T_c$:

$$T_c = \frac{eV_w}{k \ln\left(A_R T_c^2 / (j - j_{i,c})\right)} \ . \tag{22}$$

Ion current density at the cathode can be obtained from energy balance in the near-cathode region, as described in section III.2, Eq. (24). However, for the sake of simplicity ion current can be neglected, as it was done in Refs. [18, 19], and constant value of $T_c$ can be used in the right-hand side of (22). Due to logarithmic dependence of the cathode temperature on the emission current density in (22), neglecting ion current in (22) should result in a very small error. For typical ion current fraction of about 20% error in the cathode temperature is about $\alpha_i(eV_w)/(kT_c) \approx 2\%$ resulting in 5% error of the heat flux to the cathode. For the same reason, a constant value can be used for the cathode temperature $T_c$ at the right-hand side of equation (22). According to simulations[7], see figure 10 therein, typical value of the cathode temperature is about 3500 K, and its variation with current density is about 15%. Note, however, that these simplifications can result in significant errors in case of extensively cooled cathode when emission current is low. Such arc is not considered in this paper; nevertheless, formula (22) without the simplifications is applicable in this case as well. Also note that the ion current cannot be neglected when considering energy transfer in the near-cathode plasma (further in this chapter) because ions transfer significant fraction of energy.

Analytical results for the voltage drop in the near-cathode layer obtained with this approach are plotted in figure 3 in comparison with results of numerical simulations for two different pressures. Note that analytical relation for cathode voltage (18) does not include the gas pressure, the ion current and effects of other parts of the arc. The results of full simulations for the cathode voltage $V_{c\,layer}$ and the analytical solution at two different pressures are very close to each other therefore proving validity of the assumptions used in the analytical model. At lower current densities voltage in the near-cathode layer is high because major portion of the heat released in the layer is spend for heating the cathode to the temperature sufficient for maintaining electron emission. At higher current densities portion of the heat required to heat up the cathode reduces, and the cathodic voltage drop decreases asymptotically to the value $V_w + 3.2(k/e)T_{e,plasma}$.

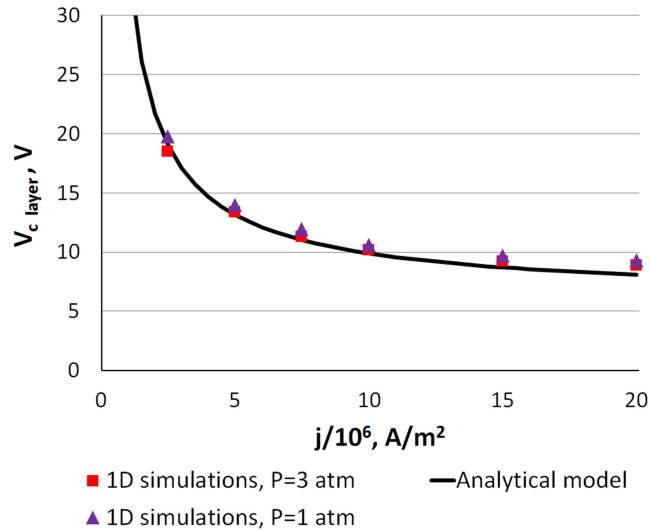

Figure 3. Voltage in the near-cathode layer as a function of total current density. Analytical solution (18) is in a good agreement with results of the simulations[7]. With decrease of the current, voltage in the layer becomes higher because larger portion of the heat released in the layer is spent for the cathode heating.

## III.2. Ion current to the cathode

The results presented above were obtained only using energy balance in the cathodic region and did not require knowledge of plasma parameters (except for electron temperature), composition of heat flux to the cathode surface and ion current. For completeness we provide description of the near-cathode plasma and determine the electron temperature and thickness of the cathode layer, which are required for coupling with the arc column model.

It is convenient to start consideration with ion current. Ion current is important mechanism of the cathode heating. As mentioned earlier, emitted electrons are accelerated in the sheath and bring their energy to plasma. Hot plasma electrons in the near-cathode layer lose their energy in inelastic collisions (excitation and ionization). The ions that impinge onto the cathode surface recombine with electrons from the cathode and therefore release ionization potential for each recombination and bring significant heat flux to the cathode. In previous analytical theoretical papers[17,19] it was assumed that the ion current is the only source of energy flux to the cathode. However, it is reasonable to assume that plasma electrons lose some portion of their energy in elastic collisions with heavy particles, and this energy is transferred to the cathode by thermal conduction (due to temperature decrease towards the cathode surface). In other words, there is some cost of ionization $\varepsilon_{ion}$ (cost of creation of a single electron-ion pair) which is higher than ionization potential:

$$q_{to\ cathode} = \varepsilon_{ion} j_{i,c} / e . \tag{23}$$

If ionization cost is known, then ion current density can be obtained from Eq. (23) using known heat flux to the cathode:

$$j_{i,c} = j \frac{e}{\varepsilon_{ion}} \left( V_w + T_c^{2.5} \frac{1}{j} \sqrt{\frac{4}{5} \frac{\lambda_c}{r_c} \sigma \varepsilon} \right). \tag{24}$$

Note that lengths of ionization and thermal non-equilibrium regions are close (see figure 1a,b), in other words, processes of ionization and elastic energy transfer from electrons to heavy particles take place in more or less the same region. Hence, conventional meaning of ionization cost should be applicable here. Ionization cost is weakly dependent on energy of electrons and pressure, and is typically about twice ionization energy for rare gases (see Ref. [30] for instance).

Comparison with results of the simulations (see figure 4) has shown that good assessment for $\varepsilon_{ion}$ is 40 eV for pressure 1 atm. and 50 eV for 3 atm. confirming that the value is rather conservative. For the sake of simplicity constant value of 40 eV can be utilized in the model of cathodic region. Assumption that all of the heat is brought to the cathode by ions corresponding to ionization cost equal to ionization energy leads to significant errors in values of ion current (significant disagreement with results of the simulations). As seen from figure 4, the ion current fraction is typically about 15%-20% and slightly decreases with total current density.

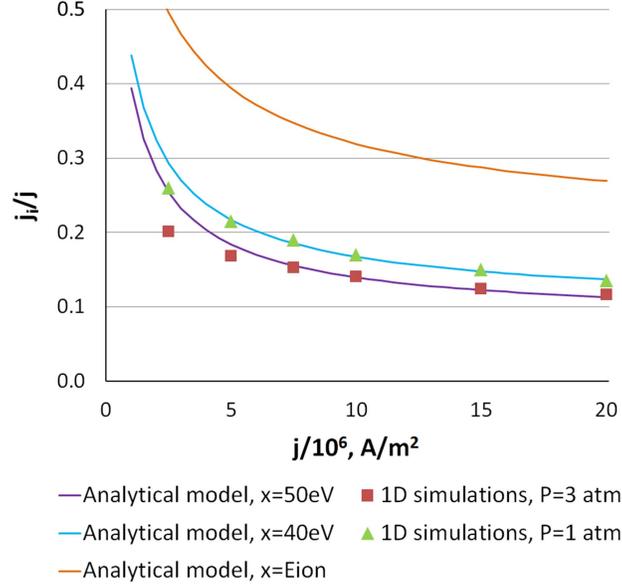

**Figure 4.** Ion current fraction at the cathode as a function of the total current density. With ionization cost $\varepsilon_{ion}$ in a range of 40 eV to 50 eV agreement between analytical solution (24) and results of the simulations[7] is obtained.

### III.3. Electron temperature in the near-cathode region

Electron temperature at the plasma edge of the cathode region, $T_{e,plasma}$, is used in relation (18) for the cathodic voltage and is needed to describe interaction of the cathodic region with the arc column. Ability to determine this parameter from the analytical arc model will make the model more self-consistent, free of heuristic approximations.

Electron temperature in the cathode region can be obtained using known ion current density to the cathode. According to the simulations, major terms in the ion transport equation (4) are diffusion $\nabla(D\nabla n)$ and source $s_i$. Hence, simplified (approximate) equation of ion transport can be written as:

$$\frac{d}{dx}\left(D\frac{dn}{dx}\right) = k_r n^3 - k_i n_a n . \qquad (25)$$

Simplified relation for the ambipolar diffusion coefficient $D$ can be used:

$$D \approx 2\frac{k(T+T_e)}{\nu_{i,a} m_{Ar}} \approx \frac{3}{8}\sqrt{\frac{\pi}{m_{Ar}}}\frac{k^{1.5}T^{0.5}(T+T_e)}{\sigma_{ia} p} . \qquad (26)$$

In expression (26), the collision frequency of a single atom with ions $\nu_{a,i}$ is neglected as compared to collisions of an ion with atoms $\nu_{i,a}$ due to rather low ionization degree in the cathode region.

The diffusion coefficient $D$, neutrals number density $n_a$ and reaction rate coefficients $k_i$, $k_r$ are temperature-dependent. For the sake of simplicity, temperature variation across the near-cathode layer is out of the scope of the paper, only level of temperature, some average value across the region is of interest. It allows to treat coefficients $D$, $k_i$, $k_r$ and $n_a$ in (25) as constants and obtain analytical solution for ion flux:

$$\frac{1}{D}\Gamma_i^2 = n^2\left(\frac{k_r}{2}n^2 - k_i n_a\right) + const, \tag{27}$$

where ion flux is $\Gamma_i = -D\,dn/dx$.

The constant in (27) can be determined from boundary conditions at the arc column side where ionization equilibrium takes place ($k_r n^2 = k_i n_a$) and ion flux is small ($\Gamma_i \approx 0$). In the vicinity of the cathode surface (at the sheath edge) plasma density is small and can be set to zero. It gives a relation for ion current density at the cathode:

$$j_{i,c} = e k_i n_a \sqrt{\frac{D}{2k_r}}. \tag{28}$$

Ion current in the left-hand side of the equation is known from previous section, Eq. (24); coefficients in the right-hand side are dependent on temperatures of the electrons and heavy particles. Note that dependence on the electron temperature is much stronger due to presence of reaction rate coefficients, especially $k_i$ (5). Hence, approximate values for the temperature of heavy particles can be utilized. In (26) it was put equal to electron temperature making right-hand side of (28) dependent on electron temperature only and allowing to express the electron temperature. Due to low ionization degree in the near-cathode region simplified relation for number density of atoms can be used: $n_a = p/(kT)$. Substitution of relations (5) into (28) and the ion current from equations (17) and (23) gives following relation for the electron temperature in the near-cathode region:

$$T_e = -\frac{T_i + 0.5T_r}{\ln\left(\dfrac{jV_w + q_{c.h.cond.}}{\varepsilon_{ion}}\dfrac{\sqrt{A_r}}{A_i}\sqrt{\dfrac{8}{3}\sqrt{\dfrac{m_{Ar}k}{\pi}}\dfrac{\sigma_{ia}}{p}\sqrt[4]{T_e}}\right)}. \tag{29}$$

Dependence of the right-hand side of Eq. (29) on electron temperature is very weak. For the sake of simplicity $T_e$ in the right-hand side of Eq. (29) can be estimated by its average value of 14 000 K giving a straight-forward relation for $T_e$. Results of application of relation (29) are plotted in figure 5 in comparison with temperature at the plasma edge of the cathode region obtained in the simulations. Because relation (29) does not take into account temperature non-uniformity in the cathode layer and gives somehow averaged temperature along the region, complete agreement between the analytical results and the simulations was not expected. However, the results appeared to be in a rather good qualitative and quantitative agreement. Therefore, one can use the electron temperature obtained from (28) as an input parameter for (16), if wants to make more accurate assessments.

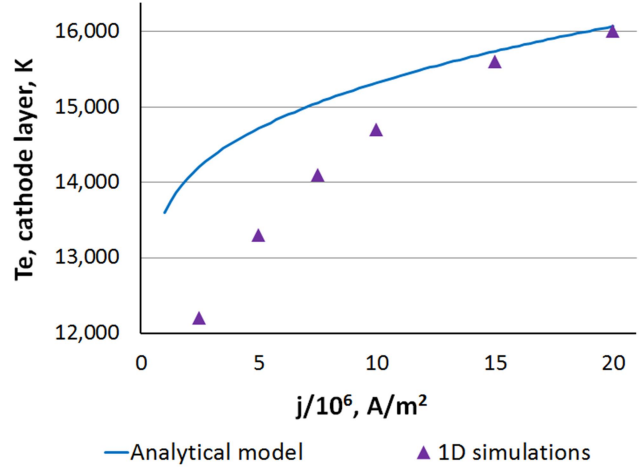

**Figure 5. Electron temperature in the near-cathode region for atmospheric pressure arc: analytical solution (29) VS simulations[7].**

### III.4. Width of the near-cathode region

Because temperature and coefficients in the equation (25) are known, it can be used for determination of the ionization non-equilibrium region width. Analytic solution of this equation can be found in Ref. [21], zero number density at the cathode surface (x=0) is assumed:

$$n_e(x) = n_{e,\infty} \tanh\left(x\sqrt{\frac{n_a k_i}{2D}}\right), \quad n_{e,\infty} = \sqrt{n_a \frac{k_i}{k_r}}. \tag{30}$$

The solution predicts asymptotic approaching equilibrium conditions. From (30), width of ionization region $L_i$ can be determined as:

$$L_i = \sqrt{\frac{D}{2n_a k_i}} ata\,nh(1-\varepsilon), \tag{31}$$

where $\varepsilon$ is tolerance (relative discrepancy from equilibrium conditions). Note that due to hyperbolic arctangents $L_i$ is weakly dependent on the tolerance value: for $\varepsilon$ of 0.5 % – 2% corresponding to large blue dots in figure 1a the length is:

$$L_i = 5\sqrt{\frac{D}{n_a k_i}} \approx \frac{5}{p}\sqrt[4]{\frac{3}{4}\sqrt{\frac{\pi}{m_{Ar}}} \frac{k^{2.5} T_e^{2.5}}{\sigma_{ia} k_i(T_e)}}. \tag{32}$$

The ratio of the length of thermal non-equilibrium region $L_T$ (see plot 1a) to the ionization non-equilibrium length $L_i$ can be assessed by ratio of relaxation lengths:

$$\frac{L_T}{L_i} \approx \frac{L_{T,relax}}{L_{i,relax}}. \tag{33}$$

Relaxation length for concentration of ions $L_{i,relax}$ is defined by a radical in relation (31). Relaxation length for the electron temperature $L_{T,relax}$ should be defined from electron heat transfer equation (8). Taking into account that at high electron temperatures near the cathode, thermal conductivity is very high and elastic/inelastic heat exchange terms are of the same order, relaxation length can be defined as:

$$L_{T,relax} = \sqrt{\frac{\lambda_e}{A^{e-H}}}. \tag{34}$$

From (33) and (34) thermal non-equilibrium width $L_T$ can be expressed:

$$\frac{L_T}{Li} \approx \frac{32\pi\varepsilon_0^2}{e^4 \ln\Lambda} \sqrt{\frac{k^{2.5}\sigma_{ia}m_{Ar}^{1.5}}{5m_e}} \sqrt{pk_r(T_e)T_e^{2.5}}. \tag{35}$$

The Coulomb logarithm was assumed to be 5. Results obtained with (32) and (35) are given in figure 6 in comparison with the results of simulations for two pressures. Rather good agreement for both pressures is observed.

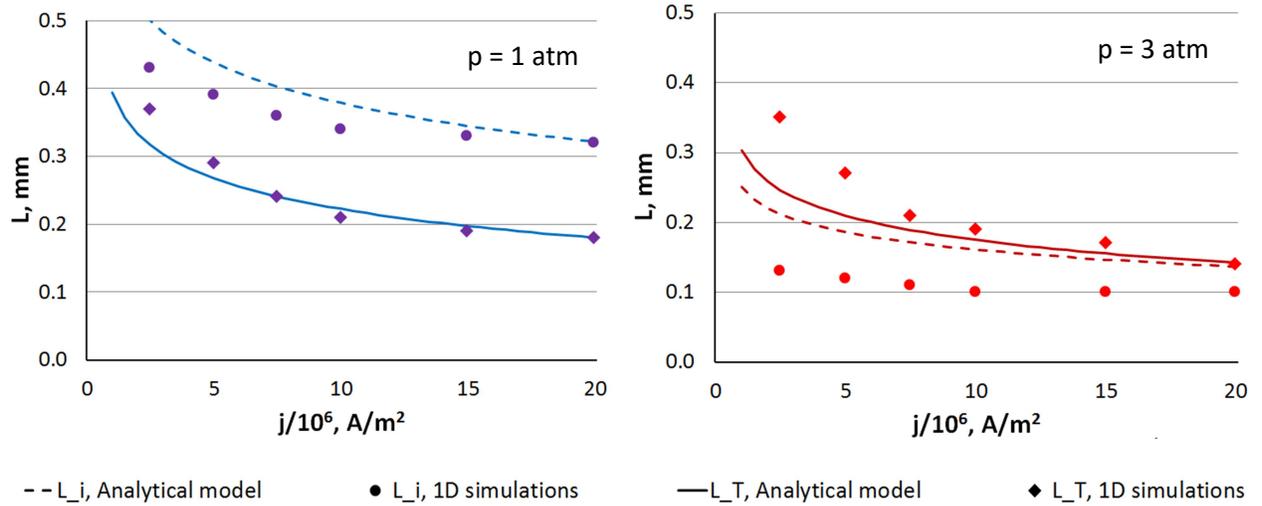

Figure 6. Width of the near-cathode ionization (35) and thermal (32) non-equilibrium layers in comparison with results of the simulations[7].

## IV. Model of the arc column

### IV.1. Description of the non-uniform equilibrium region with one equation

As previously mentioned, the arc plasma can be described with non-linear differential equations for the transport of ions (4), the heat transfer of electrons (8) and heavy particles (9), the electric field (11). Arc column is defined as a region where the thermal and ionization equilibriums are maintained: $n_e = n_{Saha}$ and $T = T_e$, whereas plasma parameters may be non-uniform. These algebraic relations can be used to reduce the number of differential equations to a single equation for one of the independent variables. It is convenient to formulate this equation for the temperature as an independent variable.

Equality of temperatures of electrons and heavy particles $(T = T_e)$ allows to write a simple relation of energy balance. Summation of equations for electron (8) and heavy particle temperatures (9) results in canceling of heat exchange term between these species and yields relation for energy balance of all plasma species as a whole:

$$-3.2 \frac{k}{e} \vec{j} \cdot \nabla T = \nabla \cdot \left( (\lambda_e + \lambda_h) \nabla T \right) + \vec{j} \cdot \vec{E} - Q^{rad}(T). \tag{36}$$

Rather similar relation for energy balance in the arc column was written in Refs. [15 and 16], however long arcs were considered in these books and convective heat transfer and radiation in the arc column were not taken into account.

Electric field in Eq. (36) can be expressed via gradient of electron density using Eq. (11):

$$1.5 \frac{k}{e} \vec{j} \cdot \nabla T + \nabla \cdot \left( (\lambda_e + \lambda_h) \nabla T \right) = \frac{k}{e} T \vec{j} \cdot \nabla \ln n - \frac{j^2}{n_e e^2} m_e \nu_{e,i} - Q^{rad}. \tag{37}$$

Note that in derivation of Eq. (37) it was taken into account that electron-ion collisions are dominant and ion flux is negligible outside the near-cathode region (last term in relation (1) was omitted). Equation (37) has only two independent variables: $n$ and $T$; transport coefficients can be expressed as their functions. In case of equilibrium, dependence between these variables is determined by algebraic relations: Saha equation (7) and equation of state (11). It allows excluding electron density from the equation (and obtaining an equation with a single independent variable – temperature).

Relation between $\nabla n$ and $\nabla T$:

$$\nabla \ln n = \nabla \ln T \left( a + b \frac{eE_{ion}}{kT} \right), \quad a = \frac{1 - 5\alpha}{4 - 2\alpha}, \quad b = \frac{1 - \alpha}{2 - \alpha}, \tag{38}$$

$\alpha = \dfrac{n}{n_a + n}$ is the ionization degree.

According to results of the simulations, the ionization degree in the arc core does not exceed 50% for current densities up to $7.5 \cdot 10^6 \, A/m^2$ (see figure 8 in the first paper[7]). Note that $eE_{ion} >> kT$ in the arc; and if $\alpha < 0.5$ then the coefficient $b$ varies in a range 0.35 – 0.5 and $|a| < 1.5b$. Therefore, the first term in the brackets in the right-hand side of (38) can be omitted and (38) can be rewritten in an approximate simplified form:

$$\nabla \ln n \approx \nabla \ln T \frac{eE_{ion}}{2kT}. \tag{39}$$

The Saha equation (7) can also be used to express the transport coefficients in (37) as functions of temperature:

$$\lambda_e = \tilde{\lambda} T^{3.5}, \quad \rho = \tilde{\rho}/T^{2.5}, \tag{40}$$

where $\rho$ is electrical resistivity and constants $\tilde{\lambda}$ and $\tilde{\rho}$ are:

$$\tilde{\lambda} = \sqrt{\frac{2\pi}{m_e}} \frac{96\pi \varepsilon_0^2 k^{3.5}}{5e^4 \Lambda_2}, \quad \tilde{\rho} = \sqrt{\frac{2m_e}{\pi}} \frac{e^3 C_{ei} E_{ion}}{48\pi \varepsilon_0^2 k^{2.5}}. \tag{41}$$

Similar relations were obtained in Ref. [31].

Substitution of relation (39) into Eq. (37) and usage of the transport coefficients (40) yields the final equation for the temperature:

$$\tilde{\lambda} \nabla \cdot \left(T^{3.5} \nabla T\right) - \left(\frac{E_{ion}}{2T} - 1.5 \frac{k}{e}\right) \vec{j} \cdot \nabla T + j^2 \frac{\tilde{\rho}}{T^{2.5}} = Q^{rad}(T) \tag{42}$$

In relation (42) it is taken into account that $\lambda_e >> \lambda_h$ and $\lambda_e \approx \lambda_{e,i}$ due to rather high ionization degree, (according to the simulations[7], $\alpha > 0.1$ in the arc column for all current densities considered).

In expressions (40), following simplified relation for the Coulomb logarithm was used. Making use of the Saha equation (7) the Coulomb logarithm can be expressed as:

$$\ln \Lambda = \Lambda_0 + \frac{eE_{ion}}{4kT}, \quad \Lambda_0 = \ln\left(\frac{8\pi \varepsilon_0^{1.5} (kT)^{9/8} h^{3/4}}{(2n_a g_i/g_a)^{1/4} (2\pi m_e)^{3/8} e^3}\right). \tag{43}$$

Note that $\Lambda_0$ is small: for temperature in the range 12 000 K – 18 000 K corresponding to the arc column (see figure 7) $\Lambda_0$ varies from –0.25 to 0.2 (for $n_a = p/(kT)$ and atmospheric pressure), whereas the second term in the right-hand side of Eq. (43) varies from 2.5 to 3.8 for this temperature range. Accordingly, simplified relation for the Coulomb logarithm in the arc core was used:

$$\ln \Lambda \approx \frac{eE_{ion}}{4kT}. \tag{44}$$

### IV.2. Two-region analytical approximation

An exact solution of the temperature equation (42) in the arc core can be obtained numerically (in 1D or 2D) using the temperature near the cathode and temperature near the anode as boundary conditions (with known widths of the near-cathode and near-anode non-equilibrium regions); values for the near-anode region can be taken from analytical model of this region described below. However, in 1D case it is possible to obtain asymptotic analytical solutions for different areas of the arc column and, therefore, describe the column with several simple relations convenient for making estimates.

In figure 7 the temperature profiles for 5 mm atmospheric pressure arc are displayed for various current densities. At high current density (starting from $5 \cdot 10^6 \, A/m^2$), the local equilibrium between the Joule heating and radiation cooling is established. It gives a nearly constant temperature in a significant part of the gap. Exclusion of the temperature gradients from equation (42) gives relation for equilibrium temperature:

$$j^2 \frac{\widetilde{\rho}}{T_{eq}^{2.5}} = Q^{rad}(T_{eq}). \tag{45}$$

Substitution of (10) and (41) into (45) yields:

$$T_{eq} = -\frac{1.69 \times 10^5 \, K}{2\ln\left(\dfrac{j}{8.5 \cdot 10^6 \, A/m^2}\right) - \ln\left(\dfrac{p}{1\,Pa}\right)}. \tag{46}$$

The right-hand side of this relation is weakly dependent on temperature, therefore, one can simply use approximate value of 15 000 K. The equilibrium temperature obtained from (46) is plotted in figure 7 with dash-dot lines and is in a good agreement with corresponding parts of temperature profiles.

Figure 7 shows that the transition region from the near cathode region to the equilibrium region of the arc column (where $T = const$) is very short. This is due to close values of temperature in the near-cathode and equilibrium regions. At the anode side situation is reversed: there is a rather long local-equilibrium part of the arc column where the thermal and ionization equilibriums persist, whereas space variations of plasma parameters are present. In this region temperature decreases when approaching the anode where the gas temperature should be equal to temperature of the electrode.

Because temperature decreases towards the anode and radiation is a strong function of temperature, radiation becomes less important and corresponding term can be omitted from equation (42). Also one can simplify the equation by taking into account that $eE_{ion}/(2kT) \gg 1.5$ and that convective heat transfer is dominating over conductive one in the most part of the arc. With these simplifications, approximate relation for temperature variation in the local-equilibrium region is given by:

$$E_{ion} T^{1.5} \vec{j} \cdot \nabla T = 2 j^2 \tilde{\rho}. \tag{47}$$

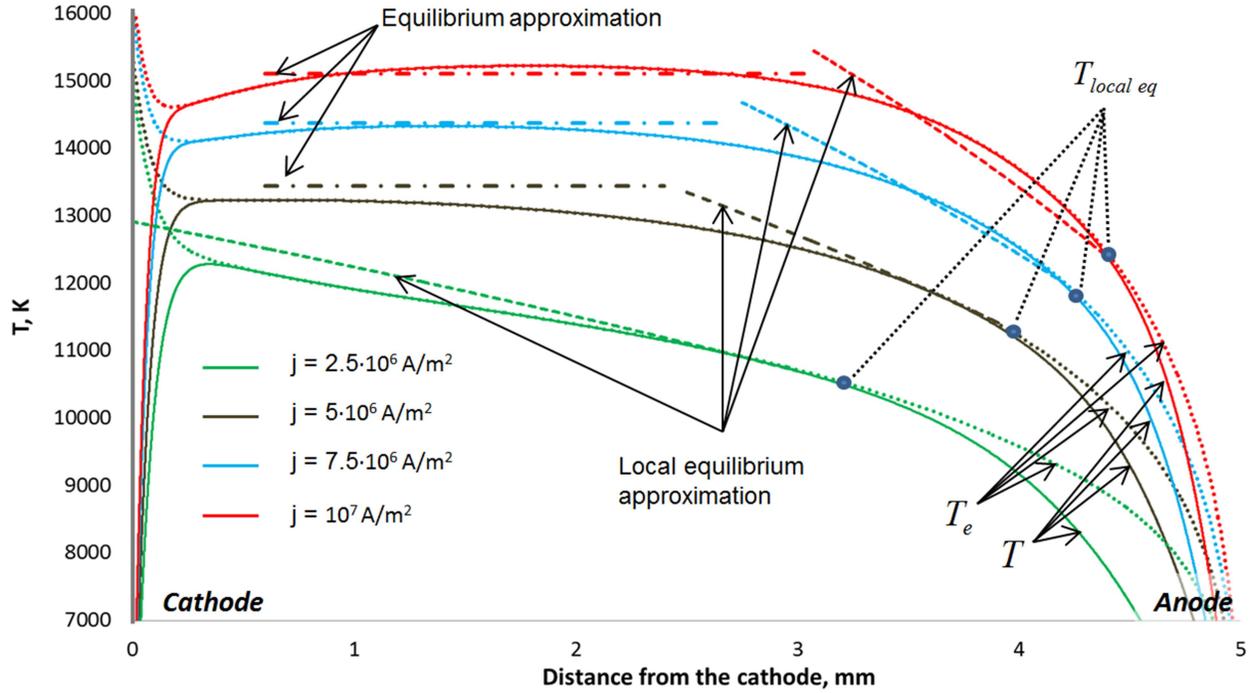

**Figure 7. Profiles of gas temperature (solid lines) and electron temperature (dotted lines) of atmospheric pressure arc at various current densities; result of simulations[7] and approximate solutions: (46) – dash-dot lines, (51) – dashed lines.**

Equation (47) has following physical meaning. As mentioned earlier, in the local equilibrium region temperature decreases towards the anode and radiation becomes small as compared to equilibrium region, resulting in low energy losses from the plasma. Conductive heat transfer in the arc column plasma is small as compared to one attributed to the convection of electrons. These simplifications allow rewriting equations for the electric field (1) and the energy balance (36) in shorter forms:

$$\vec{E} = -1.7 \frac{k}{e} \nabla T - \frac{k}{e} T \frac{\nabla n}{n} + \frac{\vec{j}}{ne} m_e \nu_{e,i}, \tag{48}$$

$$\vec{E} = -3.2 \frac{k}{e} \nabla T. \tag{49}$$

Taking into account that in the plasma column, where the Saha equation (7) is satisfied, gradients of the temperature are much smaller than gradients of the plasma density (see Eq. (39)), from equations (48),

(49) it is clear that the electric field is small as compared to its two last components in the right hand side of (48). In other words: electric field component representing electrical resistivity of plasma is almost completely compensated by oppositely directed field component representing electron diffusion caused by electron density gradient. It means that electric field can be neglected and electron flux is driven only by diffusion:

$$\vec{j} = \frac{k\,n\,T}{m_e \nu_{e,i}} \left( \frac{\nabla n}{n} - 1.5 \frac{\nabla T}{T} \right). \tag{50}$$

Note that such description of plasma behavior is not common for analytical models of near-electrode regions. Typically in such models electric field is high and is defined by plasma density gradient (second term in the right hand side of Eq. (48) i.e., satisfies the Boltzmann distribution). The present study reveals that such models can be applicable only to non-equilibrium regions significantly closer to the electrodes than non-uniform parts of the arc column, i.e. to the regions where the thermal conductivity plays important role and net ionization or recombination takes place.

First order differential equation (47) in 1D case requires only one boundary condition. Solution of this equation is given by:

$$T = \left( \frac{5 j \widetilde{\rho}}{E_{ion}} (x_{local\ eq} - x) + T_{local\ eq}^{2.5} \right)^{2/5}. \tag{51}$$

Here, $x_{local\ eq}$ corresponds to an edge of the arc column at the anode side, i.e. a location where the thermal equilibrium breaks; and $T_{local\ eq}$ is temperature value at this location. Methods of these parameters evaluation are described in further sections of the paper.

The solution (51) is displayed in figure 7 with colored dashed lines. Rather good agreement with the results of 1D simulations is observed at lower current densities. For higher current densities, the thermal conductivity starts playing noticeable role. However the agreement is still reasonable and 2-region approximation (uniform region described by (46) and non-uniform local-equilibrium region described by (51)) can be used for description of the arc column, in particular, to obtain voltage.

Voltage in the arc column according to the 2-region approximation is given by:

$$V_{col} = \frac{Q^{rad}(T)}{j} L_{eq} + 3.2 \frac{k}{e} (T_{eq} - T_{local\ eq}), \tag{52}$$

Where the first term in the right-hand side part of the equation represents voltage in the equilibrium (uniform) region, the second term stands for voltage in the local equilibrium region which was derived from relation (36), where radiation and thermal conductivity where neglected.

Using relation (51) one can determine length of the equilibrium and local-equilibrium regions:

$$L_{local\ eq} = \min\left(E_{ion} \frac{T_{eq}^{2.5} - T_{local\ eq}^{2.5}}{5j\tilde{\rho}}, L_{column}\right), \quad L_{eq} = L_{column} - L_{local\ eq}, \quad (53)$$

where $L_{column} = L_{arc} - L_{c\ layer} - L_{a\ layer}$, $L_{c\ layer}$ is defined in (32) and $L_{a\ layer}$ is defined in (85).

### IV.3. Transition to the near-anode non-equilibrium region

When approaching the anode, the temperature of heavy particles decreases (see figure 7) and becomes equal to the anode temperature at its surface. Electron temperature decrases as well due to energy exchange with the heavy particles via elastic collisions (see term $Q^{e-h}$ in Eqs. (8) and (9)). With the decrease of the electron temperature, plasma density also becomes lower. As the result, the electron-heavy particle heat exchange becomes low, and eventually deviation between temperatiures of the electrons and heavy particles takes place; notations $x_{local\ eq}$ and $T_{local\ eq}$ are used in this paper to determine this location and corresponding temperature. Frequency of inelastic collisions also significantly decreases towards the anode manifesting in rate reduction of ionization and recombination and resulting in eventual deviation from the ionization equilibrium (plasma density is no longer determined by Saha equation (7)).

In a general case, thermal and ionization equilibriums can take place at different locations. However, as the results of 1D simulations of argon arc have shown, these locations are actually very close to each other (see figures 7 and 8). For the sake of clarity we will define near-anode non-equilibrium region as an area where the ionization equilibrium breaks, because it is convenient for further considerations of the near-anode region.

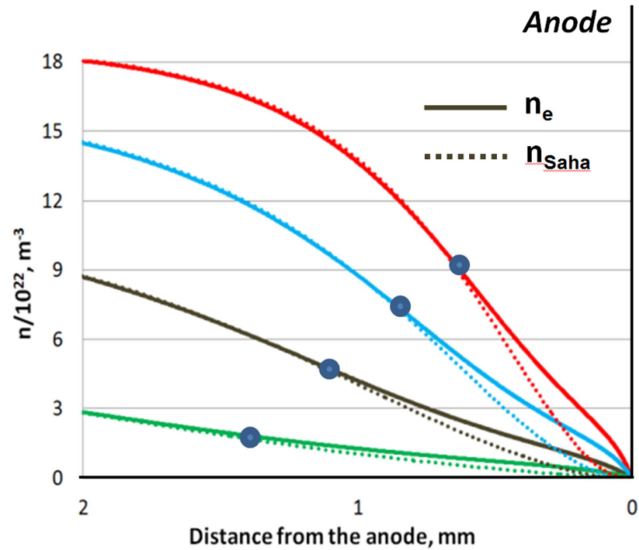

**Figure 8.** Electron number density profiles near the anode. Boundaries of the near-anode ionization non-equilibrium region are marked with circles. Color notations are the same as in figure 7.

Plasma temperature $T_{local\ eq}$ at the location where the local equilibrium breaks (transition point between the arc column and the near-anode non-equilibrium layer) can be determined making use the knowledge of variations of temperature and plasma density in the arc column; location $x_{local\ eq}$ is determined in the further section.

In the arc column, equilibrium ion number density satisfies the Saha equation and can be obtained from equation (6) which implies zero net volumetric production of ions. In fact, there is no absolute (pure) ionization equilibrium in any part of the arc: ionization and recombination reactions take place throughout the arc with more or less different rates. The difference between the ion production and their recombination is balanced by ambipolar diffusion. One can define ionization equilibrium as a state at which relative difference between ionization and recombination rates is smaller than some tolerance $\varepsilon$:

$$\frac{|k_i n_a n - k_r n^3|}{k_i n_a n} \leq \varepsilon . \tag{54}$$

As temperature decreases when approaching the anode, the ionization/recombination reaction rates become smaller, diffusion plays more significant role and the equilibrium eventually breaks. Difference between ionization and recombination rates is equal to divergence of the ion flux, so inequality (54) can be reformulated as:

$$\frac{|\nabla \vec{\Gamma}_i|}{k_i n_a n} \leq \varepsilon . \tag{55}$$

The ion flux can be determined using equation (2), which can be simplified. First of all, note that the coefficient $A_e$ is really small:

$$A_e < \frac{\nu_{e,a} m_e}{\nu_{i,a} m_a} \approx \sqrt{\frac{m_e}{m_{Ar}}} \frac{\sigma_{ei}}{\sigma_{ia}} \approx 10^{-4}$$

($\sigma_{ea}$ is about $3 \cdot 10^{-20} m^2$ and $\sigma_{ia}$ is about $10^{-18} m^2$).

The ion current is typically about several percent of total current or less, so the last term in equation (2) can be omitted. Then taking into account that $\nu_{e,a} \ll \nu_{i,a}$, $T = T_e$ and rather low ionization degree near the anode one can write approximate relation for ion current as:

$$\vec{\Gamma}_i \approx -\frac{2kT\nabla n + 2kn\nabla T}{0.5\nu_{i,a} m_{Ar}} . \tag{56}$$

Saha relation (7), (39) and temperature variation equation (47) can be used when approaching location of the equilibrium breakdown from the arc column side. The ion flux can be expressed as a function of temperature:

$$\Gamma_i \approx -2kn\frac{dT}{dx}\frac{\frac{eE_{ion}}{2kT}+1}{0.5 m_{Ar} v_{i,a}} \approx 4\frac{j\tilde{\rho}e}{m_{Ar} v_{i,a}}\frac{n}{T^{2.5}}. \qquad (57)$$

In formulation (57) $x$-axis and positive ion flux are directed from the cathode to the anode. Substitution of (57) into (55) and taking into account that $n_a \approx p/(kT)$ in $v_{i,a}$ allows to rewrite the left-hand side of local-equilibrium criterion as a function of temperature:

$$\sqrt{\frac{\pi k}{m_{Ar}}}\frac{3(j\tilde{\rho}e)^2}{4\sigma_{ia} p^2}\frac{1}{k_i(T)T^{4.5}} \leq \varepsilon \quad \text{or} \quad \sqrt{\frac{\pi k}{m_{Ar}}}\frac{3(j\tilde{\rho}e)^2}{4\sigma_{ia} p^2}\frac{1}{k_i(T_{local\,eq})T_{local\,eq}^{4.5}} = \varepsilon. \qquad (58)$$

For some fixed tolerance (it is convenient to take $\varepsilon = 0.1$ which corresponds to distinguishable difference between $n_e$ and $n_{Saha}$ on the plot in figure 8), from (58) one can determine the temperature at which ionization equilibrium breaks, $T_{local\,eq}$. In figure 9 this temperature is given in comparison with results of simulations. The temperature is plotted against current density for two different pressures. Good qualitative and quantitative agreement between the analytical model and the simulations is observed.

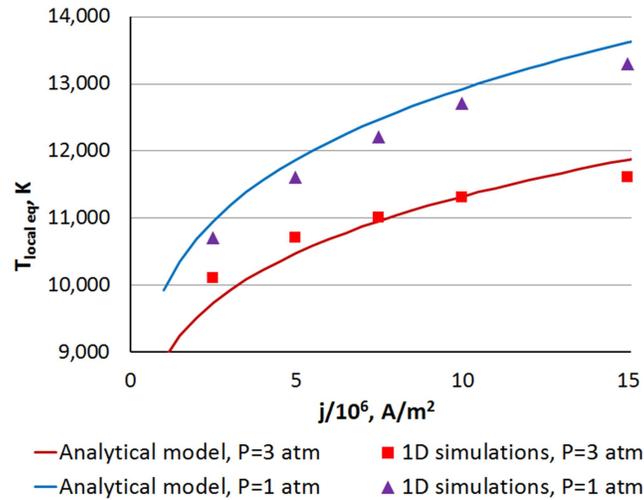

**Figure 9.** Temperature at the edge of the anode non-equilibrium layer (58) in comparison with results of the simulations[7].

# V. Model of the anodic region

## V.1. Qualitative description of the near-anode region and its structure

In the near-anode region, the electron temperature deviates from the temperature of heavy particles (see figure 10b) and plasma density deviates from equilibrium values predicted by the Saha equation (7) (see figure 10a) resulting in net recombination of ions. The ion recombination leads to significant reduction of the ion current towards the anode (see figure 10c). Similar description of the near-anode region can be found in Ref. [14], for instance.

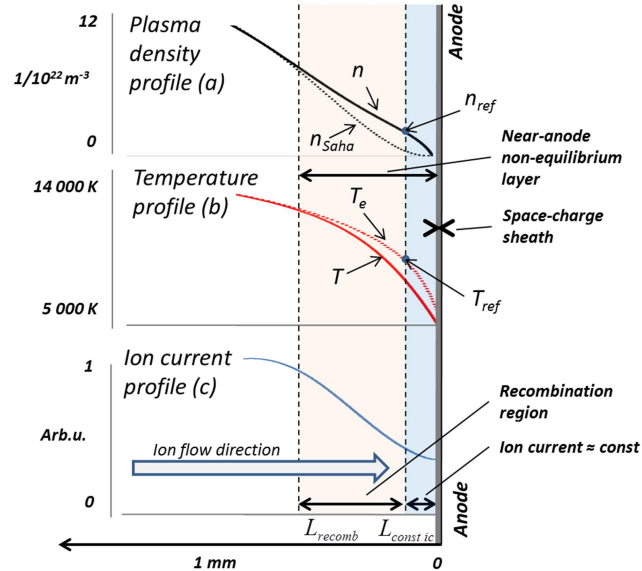

**Figure 10.** Structure of the near-anode non-equilibrium layer. Plasma density (a), temperature (b) and ion current (c) profiles obtained in the simulations[7] for current density $5\cdot 10^6$ A/m$^2$ and pressure 1 atm. are plotted.

Differential equation for conservation of ions $\nabla \Gamma_i = s_i$ or (4) should be used instead of the Saha equation (7) when describing the non-equilibrium region, making it hard to obtain analytical solution for non-equilibrium plasma in a general case. However, as can be seen from figure 10, deviation from equilibrium grows gradually towards the anode allowing to derive an analytical solution for significant part of the near-anode region using approximation of low deviation from the ionization equilibrium. Near the anode surface, the temperature of heavy particles becomes equal to the anode temperature. Despite the deviation, the electron temperature also reduces to significantly low values: simulations predict the electron temperature of about 5 500 K at the anode boundary (see figure 10b), with a weak dependence on the current density. At such low temperatures assumption of low deviation from the ionization equilibrium is apparently no longer valid. However, the ionization and recombination rates are negligible at these conditions allowing to use constant ion current relation and to obtain analytical solution.

According to the picture described above, the whole near-anode region can be analytically described using separate models for the following sub-regions: (i) recombination region where deviation from Saha equilibrium is relatively low, (ii) constant ion current region and (iii) the space-charge sheath. Corresponding models for these regions are given in the subsections below.

### V.2. Model of the recombination region

Some of the simplifications used for the local-equilibrium region of the arc column can also be applied to the recombination region of the near-anode layer: thermal conductivity and deviation between temperatures of electrons and heavy particles can be neglected. The radiation is assumed negligible, electron-ion collisions dominate over electron-atom collisions. However, because the ionization equilibrium is not maintained, approximation (44) for the Coulomb logarithm is no longer valid; 1D calculations show that constant value $\ln \Lambda = 4$ is a better approximation in this case. With these simplifications, equation (37) transforms to a simple relation between temperature and plasma density:

$$\frac{dn}{ndx} = 1.5 \frac{dT}{Tdx} + \frac{A}{T^{2.5}}, \tag{59}$$

where $A = \sqrt{\frac{m_e}{8\pi}} \frac{jC_{ei} e^3 \ln \Lambda}{3\pi\varepsilon_0^2 k^{2.5}}$ is constant (at given current density), $C_{ei} = 0.506$. Because deviation between electron and gas temperatures in the near-anode region is rather small (see Fig. 10), single temperature approximation is used in equation (59) and further in this section.

Equation (59) describes diffusion of electrons in the media featuring electrical resistivity.

Relation for the ion flux (56) can be reformulated as:

$$\Gamma_i \approx -BT^{1.5} \frac{dn}{dx}\left(1 + \frac{d\ln T/dx}{d\ln n/dx}\right), \tag{60}$$

where $B = \sqrt{\frac{\pi}{m_{Ar}}} \frac{3k^{1.5}}{4\sigma_{ia} p}$ is constant (at given pressure).

Extracting plasma density derivative from (59) and substituting it into (60) yields a relation for the ion flux:

$$\Gamma_i \approx AB \frac{n}{T} \frac{1+b}{1-1.5b}, \tag{61}$$

where $b = \frac{d\ln T/dx}{d\ln n/dx}$ – represents ratio of the thermal diffusion and conventional diffusion of electrons.

Substitution of the ion flux (61) into the continuity equation $\nabla \Gamma_i = s_i$ gives:

$$A^2 B \frac{n_e}{T^{3.5}} f = k_r n^3 - k_i n_a n, \tag{62}$$

where $f = (1-1.5b)^{-2}\left(1-b^2 + \frac{2.5}{1-1.5b}\frac{db/dx}{d\ln n/dx}\right)$. (63)

Equations (59) and (62) form a system of two differential equations with two independent variables $n_e$ and $T$. Analytical solution of these equations can be obtained making use an approximation of slow deviation from the ionization equilibrium. At the ionization equilibrium, according to (39), parameter $b$ defined in (61) is about $eE_{ion}/(2kT)$, i.e. is small (about 0.15), therefore function $f$ defined in (63) can be approximated by unity. When gradually departing from the ionization equilibrium, parameter $b$ is expected to gradually increase, nevertheless, for some part of the non-equilibrium layer approximation of small $b$ and of function $f$ equal to unity should be sufficiently accurate.

A relation for the degree of deviation from ionization equilibrium can be obtained from equation (61) making use of the relation (6) for the equilibrium density:

$$\left(\frac{n}{n_{Saha}}\right)^2 = 1+\gamma, \tag{64}$$

where $\gamma = \dfrac{A^2 B k f}{k_i(T) p T^{2.5}}$.

Substitution of a relation (64) for the plasma density into the definition of $b$ yields a relation for the parameter $b$ as a function of temperature:

$$b = 2\left(\frac{1}{2} + \frac{eE_{ion}}{kT} + \frac{1}{1+1/\gamma}\left(\frac{d\ln f/dx}{d\ln T/dx} - 2.5 - \frac{d\ln k_i(T)}{d\ln T}\right)\right)^{-1}. \tag{65}$$

When close to the ionization equilibrium, ionization coefficient $k_i$ is high and, according to (58), $\gamma \ll 1$. With decrease of the temperature, the ionization coefficient becomes exponentially low and eventually the parameter $\gamma$ increases to unity and more. Define this temperature as reference temperature $T_{ref}$:

$$\frac{A^2 B k f}{k_i(T_{ref}) p T_{ref}^{2.5}} = 1. \tag{66}$$

At this temperature ionization and recombination rates are already very low and ion current does not change, therefore, reference temperature $T_{ref}$ can be used to determine a boundary between the recombination region and the region of constant ion current. Making use of the Arrhenius expression for ionization coefficient (5), reference temperature can be approximated as:

$$T_{ref} = \frac{T_i}{\ln\left(\frac{A_i p T_{ref}^{2.5}}{A^2 B k}\right)}, \qquad (67)$$

where $T_i$ and $A_i$ are the Arrhenius coefficients (5). Values of the reference temperature change in a range of about 9 000 K – 10 000 K for various current densities. For the sake of simplicity temperature in the right-hand side of relation (67) can be put equal to 10 000 K with no significant effect on the accuracy.

Approximate relation for the parameter *b* obtained using the Arrhenius expression for the ionization coefficient reads:

$$b \approx \frac{2kT}{eE_{ion} - kT_i/(1+1/\gamma)}. \qquad (68)$$

Value of the parameter *b* corresponding to the reference temperature is given by:

$$b_{ref} = \frac{4kT_{ref}}{2eE_{ion} - kT_i}. \qquad (69)$$

As can be seen from relation (69), parameter *b* is still rather low at the edge of the recombination region (it is about 0.25). It justifies accuracy of the formula obtained taking function $f$ defined in (63) equal to unity for the whole recombination region. To illustrate this statement, in figures 11a and 11b the degree of deviation from the ionization equilibrium and parameter *b* are plotted as functions of temperature. Analytical formulae (equation (64) and (68) with $f \equiv 1$) are compared to the results of simulations; when plotting results of simulations, electron temperature was used for x-axis. As can be seen from figure 11b, parameter *b* is below 0.25 in a temperature range between $T_{ref}$ and $T_{local\,eq.}$ corresponding to the recombination region. Plasma density predicted by formula (64) (figure 11a) is in a very good agreement with the results of simulations in the recombination region. Interestingly, even at lower temperatures down to 8 000 K the agreement is still rather good. Actual plasma density is higher than equilibrium one (meaning that net recombination takes place). Note that the results potted in figure 11 were obtained for some arbitrary chosen current density (5·10$^6$ A/m$^2$), nevertheless for other current densities the plots are qualitatively the same.

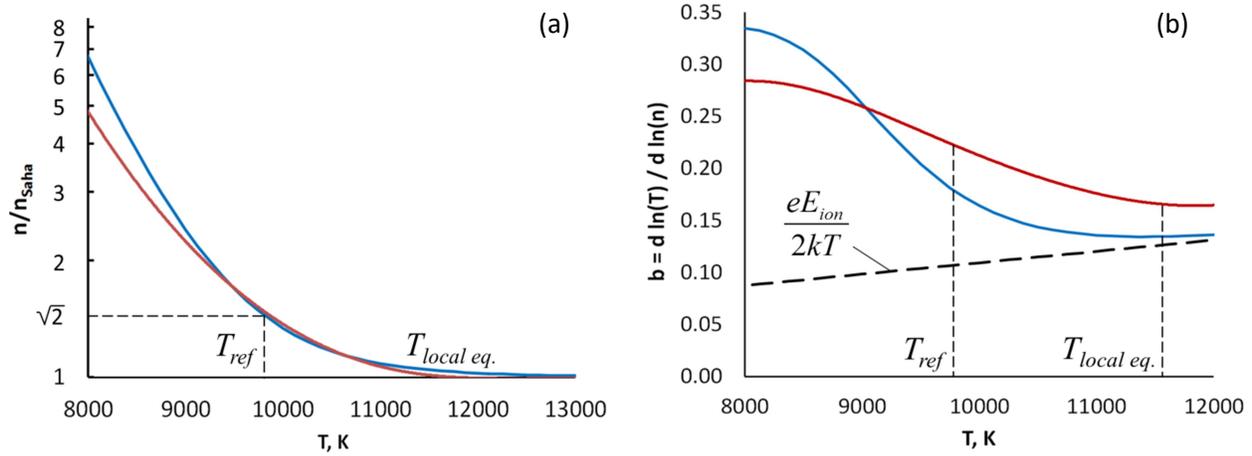

**Figure 11.** Deviation from ionization equilibrium as a function of temperature for current density 5·10⁶ A/m² and pressure 1 atm.: (a) degree of deviation; (b) ratio of logarithms. Results of the simulations[7] – blue lines, analytical solution – red line.

In figure 12 the ion current density given by the analytical solution (equation (61) with plasma density and parameter *b* defined by (64) and (68) respectively) is plotted as a function of temperature and compared to the results of simulations. Good agreement is observed in the temperature range of interest corresponding to the whole recombination region, up to the boundary with the constant ion current region. As one of conclusions, relation (61) can also be used to predict the ion current at the constant current region and the ion current to at the anode surface, if plasma parameters at the edge of the recombination region (where $T = T_{ref}$) are taken:

$$\Gamma_{i,a} \approx AB \frac{n_{ref}}{T_{ref}} \frac{2eE_{ion} - kT_i + 4kT_{ref}}{2eE_{ion} - kT_i - 6kT_{ref}}, \tag{70}$$

where $n_{ref} = \sqrt{2} n_{Saha}(T_{ref})$ is plasma density at the edge of the recombination region.

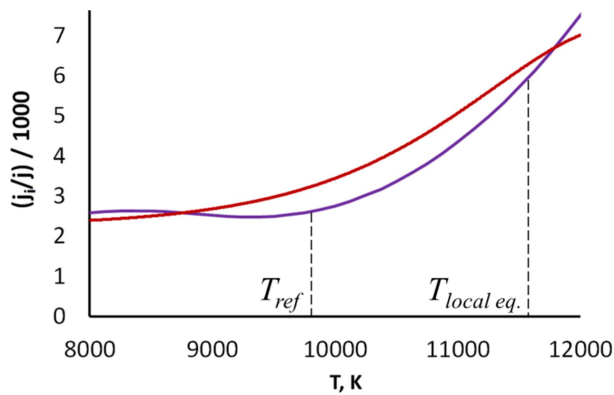

**Figure 12.** Fraction of the ion current density as a function of temperature for atmospheric pressure arc; the total current density is 5·10⁶ A/m². The results of simulations[7] – red line, analytical solution (61) – blue line.

Ion current density at the anode predicted by the analytical model (70) is plotted in figure 13 as a function of arc current density for two different pressures and compared to results of the 1D simulations. The ion current to the anode significantly decreases with increase of pressure. For both pressures considered analytical and numerical the results are in a quite good agreement.

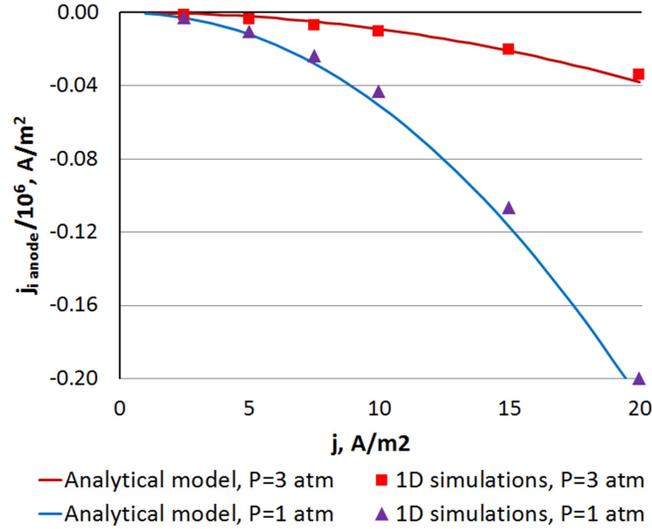

**Figure 13. Ion current density (70) at the anode as a function of the total current density.**

Knowledge of the plasma parameters at both sides of the recombination region allows to approximate its thickness and voltage. Thickness of the recombination region can be estimated from integration of equation (59) over the region:

$$L_{recomb} = \frac{1}{A} \int_{T_{ref}}^{T_{local\ eq}} \left(\frac{1}{b} - 1.5\right) T^{1.5} dT . \qquad (71)$$

Making use of relation (68) for parameter $b$ with a definition of the reference temperature (66), the integral in (71) can be approximated:

$$L_{recomb} \approx \frac{1}{2A} \left[ \left(\frac{eE_{ion}T_{av}^{0.5}}{k} - 3T_{av}^{1.5}\right) \Delta T + \frac{T_{ref}^{2.5}}{2} \right]. \qquad (72)$$

Here, $T_{av} = (T_{local\ eq} + T_{ref})/2$ and $\Delta T = T_{local\ eq} - T_{ref}$.

Voltage drop in the recombination region can be determined using similar relation as for local equilibrium region of the arc column (52), because similar assumptions were employed for these regions:

$$V_{recomb} = 3.2 \frac{k}{e} \left(T_{local\ eq} - T_{ref}\right). \qquad (73)$$

## V.3. Model of the anode space-charge sheath and heat transfer in the anode

It is convenient to describe the anode space charge sheath before describing constant ion current region, because the near-anode plasma density determined in this section will be used in the model of constant ion current region described in the next section.

Temperature of the anode surface can be evaluated using equation (15) at known heat flux to the anode. The latest can be determined from equation (31) of Ref. [7]. Because ion current to the anode is a very small fraction of total current, this equation can be simplified:

$$q_{to\,anode} = j(V_w + \max(V_{sh,a,0})) + 2.5\frac{k}{e}(jT_{e,anode} + j_e^{emiss}(T_{anode} - T_{e,anode})). \tag{74}$$

In this equation, major contributions are $jV_w$ (work function is 4.5 V) and $2.5jT_{e,anode}$ ($T_{e,anode}$ is about 0.5 eV). Sheath voltage is typically of order of 0.5 V. The last term in the second brackets is usually small and can be omitted because either emission current is small or the electron temperature becomes close to the anode temperature. The sheath voltage is typically small as compared to work function. Making use of these simplifications, substitution of (74) into (15) yields a relation for the anode surface temperature:

$$T_{anode} = \left(j(V_w + 1.25V)\sqrt{\frac{5}{4}\frac{r_a}{\lambda_a \sigma \varepsilon}}\right)^{0.4}, \tag{75}$$

where $\lambda_a$ is thermal conductivity of the anode material, $r_a$ is the anode radius. Dependence of the anode temperature on the current density was plotted in Fig. 10 of Ref. [7].

With known anode temperature, the ion current to the anode $j_{i,a}$ determined in the previous section and the total current density (input parameter of the problem), anode sheath voltage drop can be determined from balance of charged species fluxes at the anode surface and collisionless sheath boundary conditions given by equations (22)-(27) of Ref. [7].

In the hot anode case, the sheath voltage drop is assumed positive: it suppresses excess of electrons emitted from the anode surface and gives positive contribution to the total arc voltage (see Fig. 13 in Ref. [7]). Note that, in case of strong electron emission, electric potential profile in a near-electrode sheath can become non-monotonic featuring potential well and resulting in suppression of both electron fluxes passing through the sheath: emitted electrons and plasma electrons (see Refs. [32, 33], for instance). Such sheath is often referred to as a "double" sheath. However, such potential profile was shown[33] to be unstable as the potential well tends to be filled by ions and disappear even in case of even very rare collisions in the sheath. These complex unsteady phenomena are not considered here; for simplicity the electric potential in the sheath is assumed monotonic allowing expressing the total current density at the surface of the hot anode as:

$$j = \frac{1}{4} e n_{anode} \sqrt{\frac{8kT_{e,anode}}{\pi m_e}} - j_R \exp\left(-\frac{eV_{sh,a}}{kT_{anode}}\right) - j_{i,a}. \qquad (76)$$

Here, the first term in the right-hand side corresponds to the flux of plasma electrons towards the anode, $n_{anode}$ is the plasma density at the anode sheath edge, the second term corresponds to the flux of emitted electrons $j_R(T_{anode})$ defined by (21) partially reflexed by the electric field in the sheath. Note that the Schottky correction voltage should be set to zero when calculating emission current $j_R$, because of the electric field direction corresponding to suppression effect on the emitted electrons.

The ion flux from plasma to the anode surface is also suppressed by the sheath voltage in this case and can be expressed as:

$$j_{i,a} = \frac{1}{4} e n_{anode} \sqrt{\frac{8kT_{anode}}{\pi m_{Ar}}} \exp\left(-\frac{eV_{sh,a}}{kT_{anode}}\right), \qquad (77)$$

where $n_{anode}$ is plasma density at the anode sheath edge. Expressing the Boltzmann exponent from Eq. (77) and substitution it into Eq. (76) yields quadratic equation for $n_{anode}$ with a solution given by:

$$n_{anode} = \frac{1}{e}\left(j + \sqrt{j^2 + 4 j_R j_{i,a} \sqrt{\frac{T_{e,anode} m_{Ar}}{T_{anode} m_e}}}\right) \sqrt{\frac{\pi m_e}{2kT_{e,anode}}}. \qquad (78)$$

In (78) it was taken into account that the ion current density at the anode is a small fraction of the total current density. Substitution of (78) into (77) yields resulting expression for the anode sheath voltage drop in the case of hot anode:

$$V_{sh,a} = \frac{kT_{anode}}{e}\left[\frac{1}{2}\ln\left(\frac{T_{anode} m_e}{T_{e,anode} m_{Ar}}\right) + \ln\left(\frac{j}{j_{i,a}} + \sqrt{\left(\frac{j}{j_{i,a}}\right)^2 + 4\frac{j_R}{j_{i,a}}\sqrt{\frac{T_{e,anode} m_{Ar}}{T_{anode} m_e}}}\right) - \ln 2\right]. \qquad (79)$$

In case of cold anode when no electron emission takes place, the sheath voltage drop is negative to suppress the electron flux from the plasma (see Fig. 13 in Ref. [7]); and balance of the charged species fluxes at the anode surface can be written as:

$$j = \frac{1}{4} n_{anode} \sqrt{\frac{8kT_{e,anode}}{\pi m_e}} \exp\left(\frac{eV_{sh,a}}{kT_{e,anode}}\right) - j_{i,a}. \qquad (80)$$

Plasma density at the anode sheath edge $n_{anode}$ can be determined from Bohm's criterion for the ion current:

$$j_{i,a} = \frac{1}{4} n_{anode} e \sqrt{\frac{k(T_{anode} + T_{e,anode})}{m_{Ar}}}.  \tag{81}$$

Substitution of the $n_{anode}$ into yield resulting expression for the anode sheath voltage drop in the case of cold anode:

$$V_{sh,a} = \frac{kT_{e,anode}}{e}\left[\frac{1}{2}\ln\left(\frac{(T_{anode}+T_{e,anode})m_e}{T_{e,anode}m_{Ar}}\right) + \ln\left(\frac{j}{j_{i,a}}\right)\right]. \tag{82}$$

The anode sheath voltage is plotted in figure 14 against current density. In the case of hot anode, the sheath voltage drop is positive to suppress high electron emission from the anode surface. The voltage drop increases with current density reaching of about 1 V at current density of $2\times10^7$ A/m$^2$. This behavior is in a qualitative agreement with anode voltage measurements[34] in a carbon arc with hot anode. Analytical solution is in a good agreement with the result of simulations. In the case of cold anode, the sheath voltage is decreasing with current density. At higher pressure of 3 atm., the sheath voltage is slightly higher than at 1 atm. in both cases of cold and hot anode. At atmospheric pressure, the sheath voltage is negative. Good qualitative agreement is obtained between the analytical model and simulations. At pressure of 3 atm., the sheath voltage is about zero (with very small absolute value) and is not plotted in figure 14. In this case it was taken equal to zero in the analytical model.

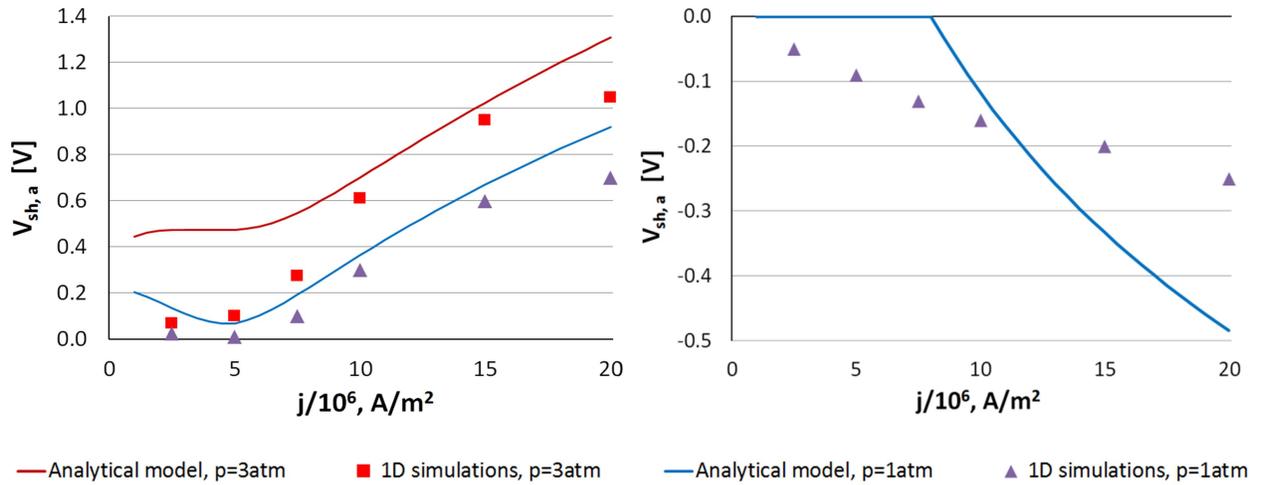

**Figure 14. Voltage drop in the anode sheath as a function of current density. Left – hot (self-cooled) anode; anal. solution is given by Eq. (79); the voltage drop is positive and increases with current density. Right – cold (1000 K) anode, anal. solution is given by Eq. (82); the voltage drop is negative and decreases with current density.**

## V.4. Model of the constant ion current region and integral characteristics of the near-anode layer

In the vicinity of the anode, plasma density decreases several orders of magnitude, i.e. relative variation of the density is much higher than one of temperature and, therefore, equations (56) and (60) can be reformulated:

$$\Gamma_{i,a} \approx -\frac{3}{4}\sqrt{\frac{\pi}{m_{Ar}}} \frac{(kT)^{1.5}}{\sigma_{ia} p} \frac{dn}{dx}. \tag{83}$$

Here, $\Gamma_{i,a}$ is constant ion current in the vicinity of the anode know from (70). Approximate integration of this relation gives thickness of the constant ion current region. Neglecting the ion number density at the anode sheath edge and taking into account that plasma density at the boundary with recombination region is $\sqrt{2} n_{Saha}(T_{ref})$ yields:

$$L_{const\ ic} \approx \sqrt{\frac{\pi}{2m_{Ar}}} \frac{3k^{1.5}(T_{ref}^{1.5} + T_{e,anode}^{1.5})}{4\sigma_{ia} p} \frac{n_{Saha}(T_{ref})}{\Gamma_{i,a}}, \tag{84}$$

where $T_{e,anode}$ is electron temperature at the anode sheath edge, which is about 5500 K according to the results of simulations, independently on arc current density.

Note that because $T_{e,ref}$ is typically about 9 000 – 10 000 K, i.e. about 1.5 times larger than $T_{e,anode}$, error in $T_{e,anode}$ should not significant affect accuracy of $L_{const\ ic}$. Moreover, it should not influence accuracy of length estimation of the whole near-anode layer because region of constant ion current is its minor part. Width of the near-anode non-equilibrium layer given by

$$L_{a,layer} = L_{recomb} + L_{const\ ic} \tag{85}$$

is plotted in figure 15 as a function of current density for two pressure values. As one can see, the layer width significantly decreases with increase of current density and slightly decreases with increase of pressure. Analytical solution (solid lines) and results of simulations (markers) are in a good agreement. According to analytical model and simulations, the length does not depend on temperature of the anode, so results for cold anode are not plotted.

In figure 15 the results of heuristic approximations non-equilibrium layer length are also plotted for comparison. As in Refs. [8, 14], non-equilibrium layer length is estimated by relaxation length of recombination processes:

$$L_{i,a}^{assess} = \sqrt{D t_r} = \sqrt{\frac{D}{\tilde{k}_r \tilde{n}^2}}, \tag{86}$$

where $D$ is the ambipolar diffusion coefficient, $t_r$ is characteristic time of recombination, $\tilde{k}_r$ and $\tilde{n}$ are characteristic values of the recombination coefficient and plasma density in the near-anode region. There is some freedom in determining parameters $\tilde{k}_r$ and $\tilde{n}_i$. In figure 15 results obtained with two different methods for evaluation of these parameters are plotted: (i) the parameters are evaluated at the edge of the non-equilibrium region $\tilde{n} = n_{Saha}(T_{local\ eq.})$, $\tilde{k}_r = k_r(T_{local\ eq.})$ (dashed line); (ii) the parameters are evaluated at the edge on recombination region $\tilde{n} = n_{ref}$, $\tilde{k}_r = k_r(T_{e,ref})$ (dash-dot line). As can be seen from the figure, using parameters at the edge of the non-equilibrium region yields much lower values of the layer thickness than both the simulations and the analytical model (85). With parameters at the edge of the recombination region, estimation (86) gives closer values but still analytical model is in much better agreement with the results of the simulations. However, with both definitions of $\tilde{k}_r$ and $\tilde{n}_i$ formula (86) gives correct trend and can be used for rough estimates of the near-anode region length.

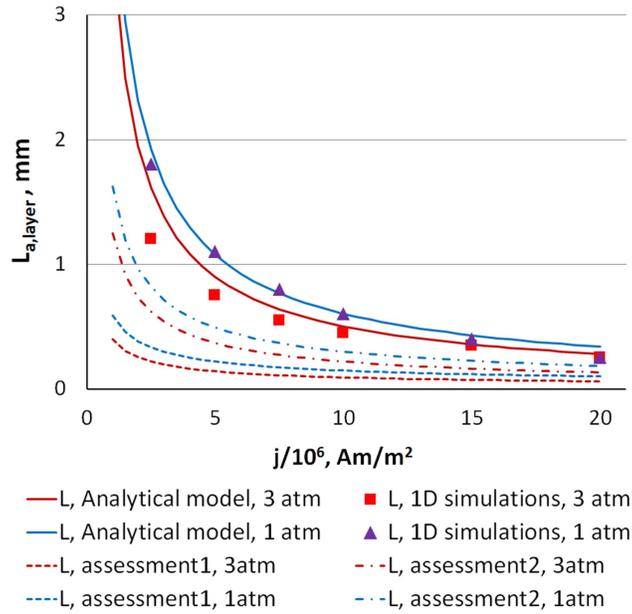

**Figure 15. Thickness of the near-anode non-equilibrium layer.**

Because the constant ion current region is significantly thinner than both the recombination region and the local-equilibrium region of the arc column, effects of thermal conductivity cannot be omitted in equation (36), and similar approach (relation (52)) for determination of voltage drop in the region would result in a significant error. More accurate approach is to use the generalized Ohm's law (1). According to the simulations, friction terms are of minor effect; therefore, they were omitted for the sake of simplicity. With this simplification, approximate integration of equation (1) over the region gives:

$$V_{const\ ic} \approx -\frac{k}{e}\frac{T_{ref}+T_{e,anode}}{2}\ln\frac{n_{ref}}{n_{anode}} - 1.6\frac{k}{e}\left(T_{ref}-T_{e,anode}\right). \tag{87}$$

The voltage drop in the near-anode layer defined as

$$V_{a,layer} = V_{recomb} + V_{const\ ic} + V_{sh,a} \qquad (88)$$

is plotted in figure 16. Two different cases were considered: cooled anode having a fixed temperature of 1000 K and a hot thermal conductive anode (of cylindrical shape). As one can see, in case of cold anode voltage drop in the layer decreases (absolute value increases) with increase of current density. Whereas in case of hot anode, voltage drop increases only at low current densities but generally positive trend is observed. Therefore, in case of cold anode, arc constriction in the near-anode layer is energetically advantageous contrary to the case of hot anode. The trends are pressure-independent: for higher pressure voltage in the near-anode layer is about 1 V higher at different anode cooling mechanisms and all current densities. Rather good agreement between analytical model and simulations is observed in case of hot anode and reasonable agreement – in case of cold anode.

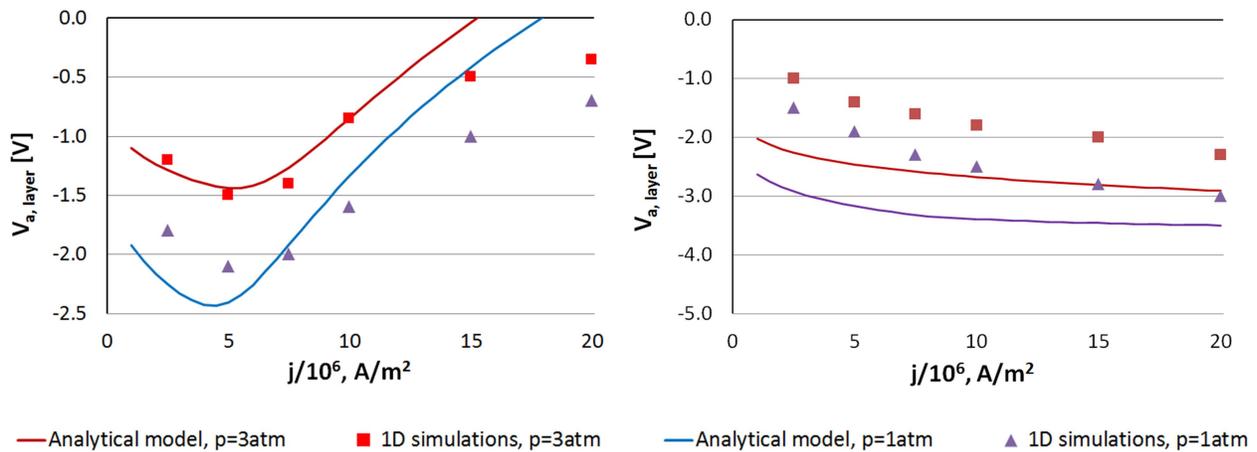

Figure 16. Voltage drop in the near-anode non-equilibrium layer. Left – hot (self-cooled) anode. Right – cold anode (1000 K).

## VI. Volt-ampere characteristic of electric arc as a whole, validation against experimental data

Arc voltage can be determined as a sum of voltages in its constituting regions:

$$V_{arc} = V_{a,layer} + V_{column} + V_{c,layer}. \qquad (89)$$

In figure 17 voltages in the whole arc (89), in the arc column (52) and in the near-cathode region (18) are plotted as a function of current density, for 2 different pressures. The hot anode arc is considered with 5 mm gap between the electrodes of 6 mm diameter. The electrodes were 10 cm long in the simulations; in analytical model infinitely long electrodes are assumed. As can be seen from the figure, at lower current densities arc voltage is primarily contributed by the cathodic region: voltage in the near-cathode region $V_{c,layer}$ and arc voltage $V_{arc}$ are close and show similar decreasing trend. The near-cathode voltage is even higher than the total voltage at current densities below $5\cdot 10^6$ A/m$^2$ meaning that contribution of negative voltage in the near-anode region is larger than contribution of arc column voltage which is short or even absent at small current densities. At higher current densities cathodic voltage continues to decrease but total voltage deviates from it and starts to increase. At current density of $2\cdot 10^7$ A/m$^2$ difference between the total voltage and the cathodic one reaches 6 – 9 V depending on pressure. This additional voltage is mostly gained in the arc column. With increase of current density, the arc column becomes longer as near-electrode layers become shorter; equilibrium region of the column heats up to have better electrical conductivity, it leads to higher energy losses by radiation included in the model and, as a result, higher electric field.

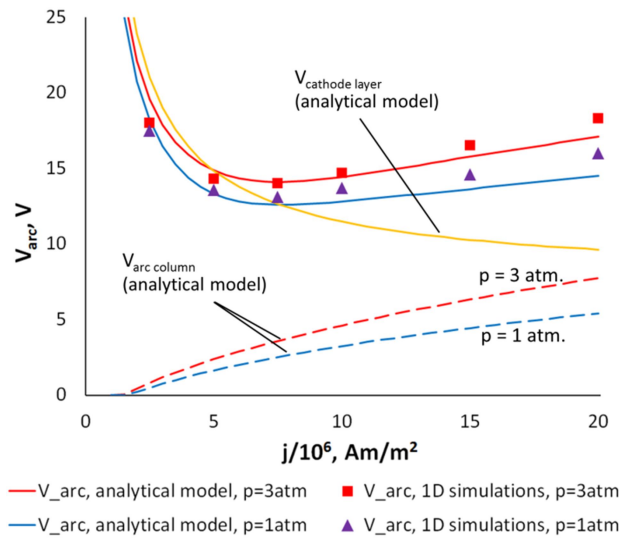

**Figure 17. Arc voltage as a function of current density for 5 mm arc.**

Analytical model of the arc was validated against experimental data[23,24]. Atmospheric pressure argon arc with cylindrical tungsten electrodes 3 mm in diameter was run at arc currents of 30 A, 50 A and 100 A. Inter-electrode gap size was varied from 0.3 mm to 3.5 mm.

In figure 18 arc voltage is plotted against current density. Reasonable qualitative and quantitative agreement between both models and experimental data is observed. At larger inter-electrode gaps, arc voltage linearly increases with gap size. This behavior can be explained by elongation of equilibrium region of the arc column. At smaller gaps (below 0.5-2mm, depending on arc current), near-anode and near-cathode non-equilibrium regions overlap and the trend is different. Analytical model cannot accurately describe such configuration; it just gives constant voltage corresponding to the arc length at which the near-electrode regions adjoin. However these constant values are still rather close to the experimental profiles.

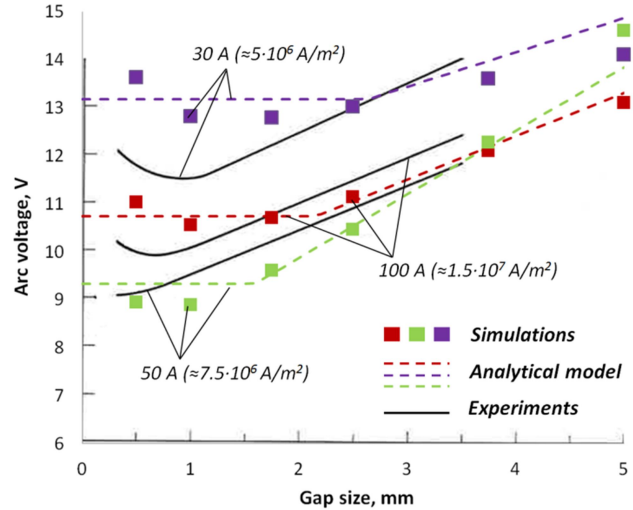

**Figure 18. Arc voltage as a function of the inter-electrode gap size for three different currents. Experiment[23] – solid lines, analytical solution (89) – dashed lines, 1D simulations[7] – squares.**

## VII. Conclusions

Self-consistent analytical model of a short atmospheric pressure argon arc comprising of models of near-electrode regions, arc column and a model of heat transfer in cylindrical electrodes was developed. Full equilibrium region and local-equilibrium region are distinguished in the arc column: in the full equilibrium region plasma parameters are uniform, in the local-equilibrium region plasma is non-uniform but the thermal and ionization equilibriums are still maintained. Near-anode region is split into recombination region and constant ion current region. The analytical model developed provides relations for following characteristics of the arc and its sub-regions.

1. Voltage drop in the near-cathode layer is given by relation (19) derived from energy balance in the cathode region. Electron temperature in Eq. (19) can be obtained from Eq. (29), or, for the sake of simplicity, constant value of 14 000 K can be utilized resulting in a fairly small error for a rather wide range of current densities ($2 \times 10^6$ A/m$^2$ to $2 \times 10^7$ A/m$^2$) and background pressures (1 atm. to 3 atm.). The cathode temperature in Eq. (19) can be obtained from Eq. (22) derived from energy balance at the cathode surface. If the cathode is not extensively cooled, ion current to the cathode can be neglected in Eq. (22) for the sake of simplicity. Otherwise, it can be obtained from Eq. (24). Width of the near-cathode region can be obtained from Eq. (32) derived from transport of ions in the region.
2. Temperature profile in the arc column can be described by the differential equation (42). The temperature profile can be constructed from two asymptotical solutions: (i) uniform profile in the equilibrium region, Eq. (46), where the Joule heating is balanced by radiation loss from the plasma, and (ii) descending profile close to the electrodes, Eq. (51), where the thermal and ionization (Saha) equilibriums are locally valid but plasma parameters vary. Temperature value at the location where thermal equilibrium breaks, Eq. (58), is used as a boundary condition.

In the non-uniform equilibrium region of the arc column, radiative energy losses from the plasma are small due to temperature decrease towards an electrode. Thermal conductivity does not play an important role, and, in accordance with the plasma energy balance, Joule heating is low. In other words, electric field can be neglected and electron flux is driven only by diffusion, see Eq. (50).

Note that typically non-uniform equilibrium region is significant only near the anode, where the electron temperature substantially decreases; near the cathode local equilibrium region is thin because temperature values are rather close in the near-cathode region and uniform part of the arc column.

3. Knowledge of the temperature profile allows obtaining the arc column voltage, see Eq. (52).
4. Voltage drop in the near-anode region and its length are obtained considering the anode region as a composition of a recombination region, constant ion current region and a space-charge sheath. An asymptotic solution for plasma density deviation from its equilibrium (Saha) value is obtained for the recombination region, Eq. (64). Temperature corresponding to the boundary of this region is determined from Eq. (67); for the region closer to the anode where the temperature is lower, recombination is negligible and approximation of a constant ion current is valid. This solution yields relations for the voltage drop in the recombination region (73), its length (72) and the ion current density to the anode (70). The latest in turn yields voltage of the constant ion current region and its length, relations (87) and (84), and sheath voltage drop, relations (79), (82) in the cases of cold and hot anode, respectively. Plasma density at the anode sheath edge in relation (87) is given by formula (78).

The analytical model was benchmarked against 1D simulations and validated against experimental data[23]. Good quantitative agreement with the results of simulations and qualitative agreement with the experimental data were obtained.

It was shown that non-equilibrium effects in the near-electrode plasma play important role in operation of the arc. When the anode is not cooled it operates at high temperatures leading to intensive electron emission and resulting in positive sheath voltage drop.

Effect of pressure variation on the lengths and voltages of the near-electrode layers and arc as a whole was investigated. It was shown that pressure variation does not affect cathodic voltage but affects anodic one: at pressure of 3 atm. the voltage is about 1 V higher than in case of atmospheric pressure; this difference is independent of anode cooling mechanisms and current density. Arc voltage is about 1 V – 2 V higher in the case of higher pressure; the effect is stronger at higher current densities when the arc column starts to play important role. The near-electrode layers become thinner with increase of pressure, especially ionization layer near the cathode, which length is inversely proportional to pressure, according to (32). Near-cathode thermal non-equilibrium layer and near-anode non-equilibrium layers are less sensitive to pressure variation.

## Acknowledgements

The authors are grateful to Vlad Vekselman (PPPL, NJ), Yevgeny Raitses (PPPL, NJ), Mikhail Shneider (Princeton University, NJ), Nelson Almeida (Universidade da Madeira, Portugal), Mikhail Benilov (Universidade da Madeira), Ken Hara (Texas A&M University, TX) and Marina Lisnyak (Université d'Orléans, France) for fruitful discussions and valuable input.

The research is funded by the U.S. Department of Energy (DOE), Office of Science, Fusion Energy Sciences.